# Isness: Using Multi-Person VR to Design Peak Mystical-Type Experiences Comparable to Psychedelics

David R. Glowacki,[1,2,3,4]* Mark D. Wonnacott,[1,3] Rachel Freire,[1,4,5] Becca R. Glowacki,[1,6] Ella M. Gale,[1,3] James E. Pike,[4] Tiu de Haan,[4] Mike Chatziapostolou,[4] Oussama Metatla[1,2]

[1]*Intangible Realities Laboratory, University of Bristol, Bristol, UK;* [2]*Dept. of Computer Science, University of Bristol, Bristol, UK;* [3]*Centre for Computational Chemistry, University of Bristol, Bristol, UK;* [4]*ArtSci International Foundation, Bristol, UK;* [5]*Rachel Freire Studio, London, UK;* [6]*Department of Design, Goldsmiths College, University of London, London, UK*

**ABSTRACT**
Studies combining psychotherapy with psychedelic drugs (ΨDs) have demonstrated positive outcomes that are often associated with ΨDs' ability to induce 'mystical-type' experiences (MTEs) – i.e., subjective experiences whose characteristics include a sense of connectedness, transcendence, and ineffability. We suggest that both ΨDs and virtual reality can be situated on a broader spectrum of psychedelic technologies. To test this hypothesis, we used concepts, methods, and analysis strategies from ΨD research to design and evaluate 'Isness', a multi-person VR journey where participants experience the collective emergence, fluctuation, and dissipation of their bodies as energetic essences. A study (N=57) analyzing participant responses to a commonly used ΨD experience questionnaire (MEQ30) indicates that Isness participants reported MTEs comparable to those reported in double-blind clinical studies after high doses of psilocybin & LSD. Within a supportive setting and conceptual framework, VR phenomenology can create the conditions for MTEs from which participants derive insight and meaning.

## 1. INTRODUCTION

In *The Doors of Perception*, Aldous Huxley recounted taking mescaline under the guidance of psychiatrist Humphrey Osmond. In a vase of flowers, Huxley reported seeing "the miracle, moment by moment, of naked existence… flowers shining with their own inner light and all but quivering under the pressure of the significance with which they were charged." He recalled how even "the folds of my grey flannel trousers were charged with is-ness."[36] The word 'psychedelic' was coined by Osmond in correspondence with Huxley in 1956. [61] Derived from the combination of the Greek words psyche (ψυχη, translated 'soul' or 'mind') and delein (δηλειν, 'to reveal', 'to make visible', or 'to manifest'), 'psychedelic' is often translated as 'mind-manifesting' or 'mind-revealed'. [12, 40] Crucially, we highlight that the word's roots are agnostic to the particular form of technology used in order to achieve 'mind-manifesting'.

In what follows, we investigate the extent to which immersive technologies (specifically multi-person VR) enable purer forms of awareness which are undistracted by ego – enabling people to tune into the 'is-ness' to which Huxley alluded. Evaluating such experiences is fraught with difficulties, because they are notoriously difficult to capture in words and metrics. To guide our efforts, we look to the resurgent field of psychedelic drug (ΨD) research, marked by Griffiths *et al.'s* influential 2006 article, *"Psilocybin can occasion mystical-type experiences having substantial and sustained personal meaning and spiritual significance"*. [22] In the intervening years, Griffiths and co-workers have accumulated evidence that ΨD efficacy in treating depression, addiction, and end-of-life anxiety correlates with their ability to occasion 'mystical-type experiences' (MTEs) which participants recount as being profoundly meaningful. [40]

Our attempts to understand whether the perceptual affordances of multi-person VR enable phenomenological experiences that create the conditions for MTEs which participants perceive as insightful and meaningful follows recent calls within human-computer-interaction (HCI) to focus on designing tools that enable experiences of meaning. For example, Mekler and Hornbaek [56] defined a conceptual framework for what constitutes an experience of 'meaning' in HCI. Three of their concepts – connectedness, resonance, and significance – have strong overlaps with concepts used to evaluate psychedelic drug experiences (ΨDEs) – namely, connectedness, ineffability, and noetic quality. [1] Light et al [50, 51] outlined concepts for technology makers to adopt in order to stimulate alternative narratives and visions, urging designers to focus on making moral progress at a time of emerging crisis and instability. They emphasize that 'significance' and 'meaning' must acknowledge human mutability and mortality within interconnected ecological systems. Kaptelinin [43, 44] has made calls for HCI to deal directly with the fundamental 'givens' of human existence (e.g., mortality, identity, isolation, meaning, etc.), in order to make our lives more 'authentic' and 'meaningful'.

*drglowacki@gmail.com

The question of meaning is important right now. With worsening climate predictions and an unprecedented rate of extinction within the biosphere, [50, 51] there is a growing sense that our every action must be balanced with awareness, including how we design and use technology. [43, 50, 51, 59] As the discourse of extinction enters into our collective psychological landscape, so does a sort of end-of-life anxiety as we struggle to shake our addiction to unsustainable growth paradigms. The ACM 'computing within limits' community has explicitly acknowledged this problem, pointing out that most computing work depends on industrial civilization's default worldview that ongoing economic growth is achievable and desirable, with a vision for the future 'very much like the present, but even more so' [59], which fails to recognize global material and ecological limits.

To date, calls for meaning-making within HCI lack empirical demonstrations showing how the proposed concepts and theoretical paradigms can be practically applied to enable experiences which participants find meaningful. In what follows, we directly address this knowledge gap. We show how immersive forms of computing can be used to cultivate awareness, ego-dissolution, and a sense of connectedness (to oneself, to others, and to the world-out-there) – all concepts with the potential to foster awareness and help us imagine our way out of the damaging and addictive paradigms in which our culture is stuck. To inspire our approach, we have turned to the ΨD research literature, because it is concerned with how to practically enable meaningful participant experiences that facilitate positive therapeutic outcomes. We outline how we have applied phenomenology from ΨD research to design the Isness multi-person VR experience, and present quantitative and qualitative evidence that Isness leads to peak experiences which occasion MTEs to which participants attribute significant personal meaning and insight, comparable to the MTEs that arise with moderate to strong ΨD doses.

## 2. BACKGROUND

**ΨD Technologies**

The 'classical ΨDs' include LSD, mescaline, psilocybin, and DMT. Phenomenologically, they produce non-ordinary and variable forms of consciousness which (compared to ordinary waking consciousness) are less centered on one's normal sense of egoic self, [8, 62] instead producing senses of unity and connectedness. [11] While the classical ΨDs were of prominent interest within psychiatry and neuroscience research in the 1950s and 60s, their recreational use and their association to counterculture prompted an end to their use in human research in the early 1970s. [61] However, the last decade has seen a resurgence in research studies carried out to evaluate their utility for promoting positive psychological health in both clinical and non-clinical settings. Double-blind trials [22, 23] and neuroimaging studies [7, 10] indicate that classical ΨDs hold strong potential as therapeutics for treating depression, addiction, and end-of-life anxiety associated with terminal illness. [40, 68] Johnson *et al.* discuss the insufficiency of the term hallucinogen' in referring to ΨDs because it suggests effects limited primarily to visual perception. [40] Given that classical ΨDs do not typically produce stark hallucinations and are instead associated with effects on human consciousness and sense of self, the term "psychedelic" has re-emerged within the scientific literature. [61]

**ΨDs and MTEs**

Early researchers identified the ability of ΨDs to facilitate powerful MTEs for participants, [28, 47, 65, 66] highlighting the correlation between subjective MTEs and the efficacy of ΨDs in treating addiction and dealing with end-of-life anxiety. The 2006 Griffiths *et al* study showed that participants who had taken psilocybin reported greater psychological well-being compared to those who had ingested methylphenidate placebo. 67% of the study participants identified their psilocybin experience amongst the most personally meaningful experiences of their lives, and analysis of their subjective reports showed that many had MTEs. These studies utilized a 'psychedelic psychotherapy' approach [15, 54], where the goal is to administer a high drug dose in order to occasion a MTE (sometimes called 'peak experience' or 'ego dissolution') and inspire subsequent behavior change (this contrasts with 'psycholytic' approaches that use lower ΨD doses). The intervening years have seen a number of additional studies, where psilocybin has been administered to healthy volunteers; [24, 26] patients with life-threatening cancer diagnoses; [25, 67] people dealing with addiction; [18, 38, 39] and those afflicted with treatment-resistant depression [13]. Several of these studies reaffirm the fact that participants' subjective reports of MTEs following ΨD ingestion offers a good predictor of positive therapeutic outcomes.

**Characterizing MTEs**

The most definitive review of features that can be used to identify a subjective experience as mystical was compiled by Stace [71] who distilled phenomenological descriptions of MTEs from a variety of sources. Building on the work of William James, [37] he identified *a sense of unity* (becoming one with all that exists) as the defining feature of the MTE. Other dimensions of MTE which Stace identified include: (1) *ineffability* (i.e., it cannot be encapsulated in words ); (2) *noetic quality* (i.e., insight into the depths of some fundamental truth or ultimate reality which transcends the discursive intellect, as captured by the Huxley quote at the beginning of this article ); (3) *sacredness* (i.e., a sense that what is encountered is holy or sacred); (4) *positive mood* (i.e., joy, ecstasy, blessedness, peace, tenderness, gentleness, tranquility, and awe); and (5) *transcendence of time and space* (i.e., conventional experiences of time and space seem to fall away).

The majority of empirical studies which have sought to measure MTEs utilize the Hood Mysticism Scale (M Scale). [35] Based on Stace's work, this scale was originally developed to measure naturally occurring (i.e., non-drug) MTEs, but Griffiths et al showed that high doses of psilocybin could reliably occasion salient MTEs in healthy participants, [22, 23, 26] and moreover that the strength of the MTE could predict positive outcomes for ΨD therapies. To specifically measure MTEs occasioned by ΨDs, Griffiths and co-workers developed and validated the Mystical Experience Questionnaire (MEQ30), [1, 52] which avoids reference to whether participants feel they have engaged with another sentience (e.g. God). It is designed to capture a participant's feeling that their experience: (1) is ineffable; (2) transcends typical experiences of space and time; (3) is mystical (i.e., produces senses of internal/external unity, connectedness, sacredness, and noetic qualities); and (4) produces a positive mood. The MEQ30 has been thoroughly tested: at the time of writing, the literature contains results from 26 previous experiments on 540 total participants, detailed in the Table 1 and 2 in the Supplemental Material (SM).

**ΨD Phenomenology and Immersive Technology**

Grof wrote that ΨDs, "used responsibly and with proper caution, would be for psychiatry what the microscope is for biology and medicine or the telescope is for astronomy… [making] it possible to study important processes that under normal circumstances are not available for direct observation" [29] It is difficult to precisely characterize the phenomenological effects whereby ΨD neurochemistry leads to MTEs, in part because ΨDs have multiple physiological effects, not all of which are likely relevant to generating altered perceptual phenomenology. A further question concerns the mechanism whereby phenomenological changes to sense perception arise during ΨDEs. For example, it has been observed that hallucinations [55] and altered perception of time [19] can arise in non-drug contexts, e.g., by placing participants in altered sensory environments. This raises an interesting question: is the 'psychedelic experience' primarily a result of 'top-down' changes in a participant's brain? Or can it also arise from 'bottom-up' changes to perceptual sensory inputs? If ΨDs offer a kind of microscope for understanding brain function, then immersive technology which achieves comparable subjective effects may have the potential to serve as a kind of microscope to unravel the subjective phenomenological threads which combine to construct MTEs.

Practically, MTEs occasioned using immersive technology might sometimes be preferable to those which arise from ingestion of ΨDs. For example, ΨDs remain subject to a host of regulatory challenges, which makes them a challenge to work with in scientific contexts. Moreover, high doses of classic ΨDs can result in an anxious, dysphoric, confusing, and (less commonly) delusional acute reaction (a "bad trip" colloquially). [40] Given that classical ΨDEs can last anywhere from 6 – 12 hours, being alert to deal with 'bad trips' in case they arise requires sustained attention from the facilitators. Finally, ingestion of classical ΨDs often leads to short-term physiological effects, including elevated blood pressure and heart rate, psychological discomfort (e.g., anxious or dysphoric reactions), and physical distress (e.g., nausea, vomiting, and headache). [40]

There are relatively few rigorous empirical studies analysing technological approaches to understand altered states. A number of VR experiences strive to give participants a glimpse of the 'trippy' visuals often associated with ΨDs (swirling geometric fractals, kaleidoscopic light trails, technicolor textures, etc., see e.g., [16]), some of which have been criticized as 'elaborate screensavers". [48] Kitson et al have outlined ways in which VR might be used to simulate the experience of lucid dreaming, [46] and Gullapalli et al have carried out studies evaluating the use of wearable technologies to *measure* drug use. [30] To date, we are aware of only one empirical study of a framework designed to *simulate* ΨD phenomenology: the 'hallucination machine' which Suzuki *et al.* designed to simulate altered visuals of the sort which participants might experience during ΨDEs. Using panoramic 360 videos (derived from Google's deep dream convolutional neural nets) which individuals could watch whilst wearing a VR headset, [73, 74] their results (N = 12) suggest that it is possible to induce visual phenomenology similar to psilocybin; however, they were unable to evoke in participants the temporal distortion commonly associated with altered states.

Our work represents a significant departure from previous approaches like that of Suzuki *et al.*: rather than simulate ΨD visuals, our focus is on how immersive technology might be used to construct MTEs comparable to those that arise during ΨDEs. Recent work by Griffiths *et al* [27] comparing MTEs that arise from ΨDs to those that arise naturally (i.e., non drug-induced) highlights the fact that MTEs *per se* are powerful predictors of lasting changes in psychological health regardless of their origins. Inspired by this insight, the remainder of this paper outlines our efforts to design and analyze *Isness*, a multi-person VR experience which we show is able to occasion MTEs similar to those which arise from large ΨD doses. The structure of this paper is as follows: First we outline the overall structure of the three-stage Isness experience, prefaced by a description of the participants and the technological components we used to build Isness. Then, we highlight specific concepts from ΨD research used to inform the design of Isness, describing how these concepts were woven into each of the 3 Isness stages. Finally, we present our results, followed by a discussion and conclusions.

**3. THE ISNESS EXPERIENCE**

**Participants**

64 healthy adults participated in Isness, where it formed part of the art installation program at a biennial psychedelics and consciousness conference held in 2019 at the University of Greenwich (London). Isness took place in two rooms on a low traffic corridor off the main conference lobby. All participants were at least 18 years old, were made aware of the potential risks associated with VR, and gave both written and verbal consent to their data being gathered and published. Over three days, Isness ran 16 times, with 64 total participants. Each group was led by one of three trained guides. To minimize participant risk, we adopted VR guidelines in line with those recommended by Madary and Metzinger. [53] The video available at vimeo.com/386402891 illustrates a few different aspects of the Isness experience, which are discussed further in what follows.

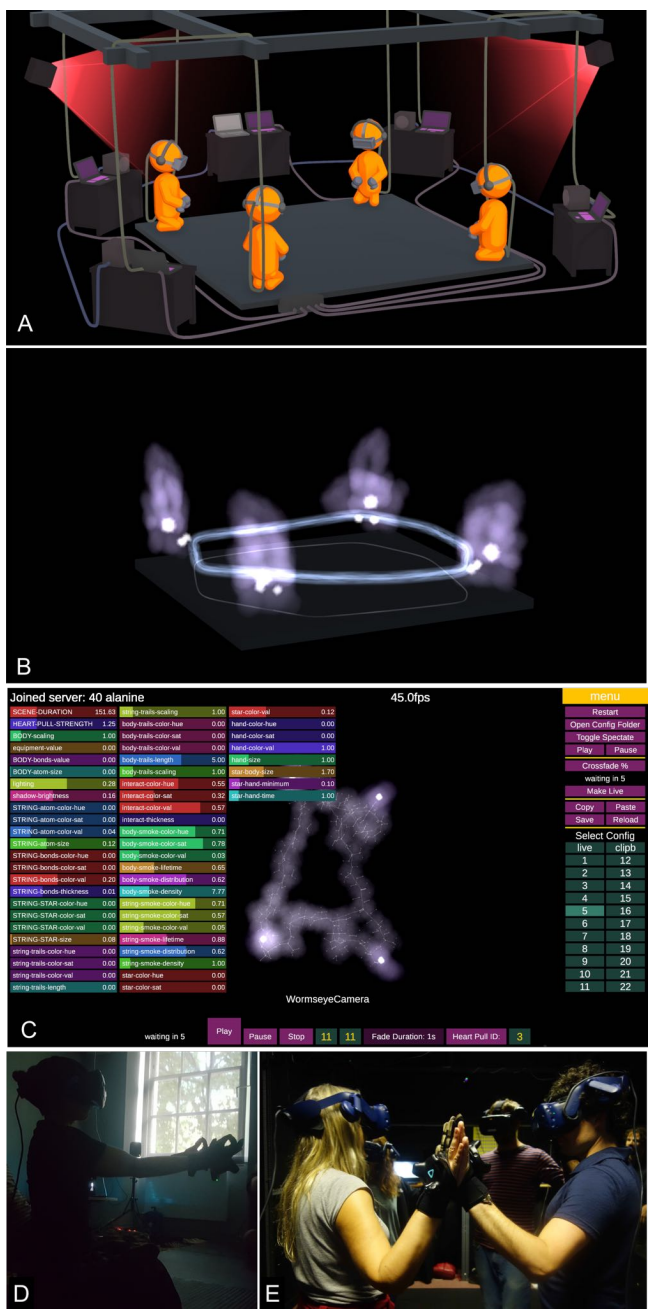

Figure 1: (A) Isness multi-person VR setup; (B) In-world view of an Isness state; (C) Interface screen for tuning aesthetic hyperparameters, showing a top-down view of 4 participants as they manipulate an energetic thread; (D) Introvertive exploration with a participant in a mudra pose; (E) Extrovertive exploration, as participants join in a circle to merge fields

### Multi-person VR setup

We developed Isness as a fork of the open-source Narupa project, [64] a multi-person room-scale VR framework (Fig 1A) [60] originally designed to enable groups of people to simultaneously cohabit real-time scientific simulation environments where they can reach out and touch molecular objects, manipulating rigorous real-time simulations of their dynamics, made possible by mounting the simulation on GPU-accelerated servers. [21, 63, 64] The client/server architecture illustrated in Fig 1A provides each VR client access to global position data of all other participants, enabling each participant to see through their headset a co-located visual representation of all the other participants (e.g, Fig 1B). We designed Isness to accommodate 4 participants wearing HTC Vive Pro headsets, each locally connected via LAN cables to the server (Fig 1A).

Using Narupa, we designed MTEs by defining a set of 'aesthetic hyperparameters', each of which controls some aspect of the participants' phenomenological experience and which can be precisely varied using the interface shown in Fig 1C. We defined a phenomenological 'state' as a given set of aesthetic hyperparameter values. The overall Isness 'journey' is comprised of a set of states, each of which has some specified time duration. This approach ensures reproducibility because it enables rigorous definition of the hyperparameter values used to design the Isness journey. The progression through the Isness journey was synchronized with a narrated soundtrack, which was played through a 4-channel sound system with one speaker mounted at each of the four corners of the VR space, as shown in Fig 1A. The complete Isness journey (comprised of 13 states) involved varying 25 different aesthetic hyperparameters, including for example: the color, distribution, density, and latency of the light bodies; the size of the heart center light; the rendering options for the energetic thread shown in Fig 1B; options for setting interactive forces to achieve different effects; the scene duration, and the global light levels. Our decisions on how to set the aesthetic hyperparameters were grounded in the design concepts discussed in section 4.

### Mudra Gloves

The 'mudra pose' plays an important role during the Isness experience. Participants adopt a 'mudra pose' by bringing the tip of their thumb in contact with the tip of either their forefinger or middle finger (Fig 1D). Within Isness, participants could use this gesture to generate light. The light-generating mudra pose was possible using custom-made 'mudra gloves' constructed using four-way stretch technical knit fabric, a sewing machine, soldering iron, and conductive fabrics. The resulting gloves fit a range of participant hand shapes, as shown in Fig 1E. The mudra glove design achieves absolute position tracking using an HTC Vive tracker mounted on a 3d printed connector attached to the glove on the back of the hand. [20] Woven into each glove is a circuit made from copper electronic textile fabric which participants can close by making a mudra pose. [20] Because the mudra gloves require no calibration and have no moving parts, they can be quickly and comfortably slipped onto users' hands without interrupting the flow of the broader Isness journey.

### Set and Setting

The $\Psi$DE is sensitive to a number of non-pharmacological contextual factors, often described as 'set' and 'setting'. [33, 34, 57, 72] 'Set' refers to the preparation of the participants, their prior psychological traits, personality structure, and their state of mind at the time of the experience. 'Setting' refers to the specific physical, social, and cultural environment in which the experience unfolds, and also the broader cultural and media discourse in which the participants are embedded. [14, 34, 78] Set and setting influence the psychological effects of any psychotropic substance (including common drugs like alcohol and nicotine), but ΨDs are particularly sensitive to these conditions. [72] A number of studies have shown that participants enter into a kind of 'hyper-suggestable state' during ΨDEs [41] – i.e., the impacts of set and setting amplify their susceptibility and responsiveness to suggestions, which have the potential to alter the contents of consciousness, magnifying whatever meaning participants bring to the experience, and influencing their perception, sensation, cognition, emotion, and behavior. [9] During Isness, we assumed that suggestibility would play a similarly important role, and therefore paid special attention to both set and setting as part of our design process. To establish a supportive set and setting, Isness was designed as a three-phase journey.

**Phase 1: Preparation**
The introductory session lasted 15 – 20 mins. Designed by a trained drama therapist, the aim was to build rapport and trust between participants and the guide who would lead their journey, mimicking the strategies used in ΨD studies to minimize adverse reactions. [22] The intro began by addressing a number of practical issues (phones off, toilet locations, placing possessions in a safe place). The guide asked participants about any health issues that might pose a risk to their participation (e.g., epilepsy, light sensitivity, medications, psychological diagnoses, communicable infections that could be spread through the VR equipment, or drugs influencing participants at present). Participants were informed what Isness would involve, made aware of potential risks (nausea, headaches, disorientation, emotional distress), and informed that they could withdraw at any point. Once they gave written and verbal consent, they took their shoes off and rubbed their hands with sanitizing lotion. The guide then explained the matter/energy framework underpinning Isness (described below), and led them through some gentle movement and breath sequences in order to draw awareness to their own embodied perception. They were invited to practice the aforementioned mudra pose, and to imagine the mudra pose as representing a coalescence between 'individual consciousness' (represented by the finger) and 'collective consciousness' (represented by the thumb). They were invited to build rapport with their fellow participants through a short exercise where they touched the palms of their hands to those of the other participants. They were then blindfolded and led by the guide into the VR room. Upon entering, they felt underfoot a soft mat. They were told that they would be safe throughout so long as they stayed on the mat. The guide then gently moved each person to a separate corner of the mat, inviting them to sit or kneel (Fig 1A).

**Phase 2: Multi-person VR Session**
Once the participants were comfortably knelt or sat at the four corners of the mat, the guide gently slipped the mudra gloves onto their hands, and initiated a 35-minute pre-recorded narrative soundtrack played on a four-channel sound system. The narrative guided participants through a short meditation, inviting them to imagine their breath as radiant light concentrated at their heart center. Each participant then removed their blindfold and was fitted with a VR headset, at which point the guide initiated the Isness VR journey, moving through 15 prespecified states, each composed from a different combination of aesthetic hyperparameters. The narrative journey was designed to balance moments of individual introvertive exploration (Fig 1D) with collective extrovertive exploration (Fig 1E). Building on evidence that ΨDs amplify the emotional impact of music, [4, 42] we accompanied the narrative by a soundtrack chosen to broadly reflect the arc of the journey.

**Phase 3: Integration**
At end of the Isness VR journey, participants were invited to go back to a kneeling or sitting pose and close their eyes. The guide removed their headsets, and they were invited to lie down, noticing what remained in their conscious awareness, and attending to their senses. They were guided on a breath meditation similar to that in Phase 1, and invited to imagine what might happen if they carried awareness of the intrinsic luminosity (they had just experienced in VR) out into the wider world – i.e., their daily reality. They were then guided to open their eyes, notice the space around them, sit up, join their palms again with their fellow participants, and make one last mudra pose. They were then greeted by their guide, who invited them to share in a 10-15 min facilitated discussion, after which they were provided a blank piece of paper for reflective writing, along with a blank MEQ30.

## 4. CONCEPTUAL FOUNDATIONS
The design of Isness is grounded in the following concepts that have been highlighted in the ΨD research literature:

**Matter as Energy**
Relationships between matter and energy often emerge in subjective accounts of ΨDEs, with material objects radiating energy and significance (e.g., Huxley's 'is-ness' quote). The Isness narrative consistently referred to the idea that matter and energy are interconvertible essences which exist on the same continuum, with participants told 'It's not just you that is made of pure energy; it's all matter, all of the molecules and atoms that interact to create all that your world contains'. The depth and rigor of this matter-energy narrative was reinforced by the fact that the molecular object which participants experienced in the Isness virtual environment as a kind of fluctuating dynamical organism (referred to at various points in the narrative as a 'molecular organism' or 'energetic thread') was *more than* some arbitrary animation; rather its dynamics are calculated in real-time using a state-of-the-art GPU-accelerated computational biophysics engine. [17] This sophistication anchored the Isness narrative in physical and scientific reality, encouraging participants to reflect on the fact that everyday material objects *are actually* constructed from the dynamical choreography of molecular organisms whose essences are fundamentally energetic. [21] The sense of continuity between energy and matter was reinforced also by dissolving the bodies of Isness participants into energetic essences (Fig 1B), and by describing the mudras as 'symbolic poses which amplify energy.' As shown in Fig 1B, the mudra light enabled participants to connect with and sculpt the dynamics of the energetic thread, and participants were asked 'how does it feel to touch pure energy?'

### Connectedness

Connectedness is an important characteristic of MTEs, [11] which emerges quite naturally from reimagining conventional matter in terms of a common energetic essence. The mudra motif was an important mechanism for facilitating a sense of connection. For example, there were a number of points during the Isness narrative where participants were encouraged to make contact between their mudras and those of the other participants (Fig 1E), in an effort to enable them to experience the energetic merging that arose in moments of touch. In a number of the Isness states, the participants' energetic bodies are rendered in a similar way as the energetic thread (Fig 1B), so that the thread is continuous with the energetic essence of the participants themselves, at which point participants were asked 'How is it to be connected to everything? To everyone?'

### Unity

Stace defined unity as a state of pure awareness uninterrupted by the brain's default tendency to construct egoic identity. [71] During introvertive (internal) unity experiences, the sense of separation of oneself and a transcendent reality is overcome, whilst during extrovertive (external) unity experiences the boundary between oneself and the world around them is dissolved. Invitations to cultivate a sense of external unity occurred during phase 2 of Isness. For example, we represented the energetic essence of each participant in the same way, anonymizing their respective identities, and implicitly encouraging them to recognize their common essence. We also encouraged a choreography whereby participants came into proximity with one another, so they could experience the fluidity of their energetic bodies merging with the other bodies in the space. Invitations to cultivate a sense of internal unity were primarily carried out in phase 1 and 3, where for example participants were guided through a short breath meditation, and encouraged to visualize their breath as light radiating from their heart center, similar to what they saw when they 'awoke' into Isness VR, where their heart centers were illuminated as shown in Fig 1B.

### Ego-dissolution

ΨDEs are characterized by a reduction in the self-referential awareness of normal waking consciousness (ego 'death', 'loss', 'disintegration', or 'dissolution'). [62] We sought to encourage a sense of ego-dissolution in two ways. First, we designed the journey around a loose arc of energetic emergence, fluctuation, and eventual dissipation, encouraging reflection on transience and 'mutability'. [50] Each Isness participant materialized in VR as three shining lights (Fig 1B): one located at the heart center, and the others at the origin of the mudra pose (Fig 1B). Over the next 25 mins, their energetic light bodies gradually intensified, leaving distinct residues as they moved through the space. Eventually their essences dissipated, leaving only blackness. Second, we recognized that the HMD acts like a kind of blindfold which replaces participants' visual sensory inputs with different inputs. [75] We sought to encourage a sense of ego-dissolution by inviting participants to focus less on their own internal ego narrative and engage in embodied forms of sensing. For example, participants were invited to 'move toward one another and form a circle, and place the palms of your hands together' (Fig 1E). Such moments of contact required tuning into non-visual senses (kinaesthetic, proprioceptive, and tactile), diverting attention away from internal ego narratives.

### Transcendence of Space and Time

VR is particularly well suited to exploring alterations in our experience of space and time. [31, 32] Isness included specific states that challenged participants' conventional understandings of space and time. For example, as they sculpted the dynamics of the energetic thread, they were invited to become aware that whilst their own bodies were subject to normal space constraints and unable to pass through the floor, the same was not true for the energetic essences cohabiting the simulated VR space with them. Over the arc of the VR journey, the states gradually evolve to a point where participants can simultaneously perceive both the past and present, and they are asked 'what is it like to see the past?'

### Noetic Quality

Noetic quality is often associated with a subjective experience of something greater than oneself. For example, encouraging participants to think about themselves, and everything around them as having a fundamental energetic essence was an invitation for them to imagine themselves as part of a larger energetic unfolding. They were reminded that 'you are simply energy in motion', a sense which was reaffirmed by the simulated dynamics of the 'energetic thread'. At various points during the Isness narrative, participants were encouraged to engage with the energetic organism, and actively sculpt its dynamics. These active moments of engagement were balanced by moments of stillness, with participants encouraged to 'Explore the feeling of both stillness and motion. Of being and doing. Notice what happens if you do nothing.' In doing nothing, participants discovered that the energetic organism carried on, following its own intrinsic choreography, creating the sense of an object with a sort of otherworldly intelligence – whose 'is-ness' is manifest as a kind of perpetual motion which follows a different logic.

## 5. ANALYSIS

Each of the 64 participants completed all three Isness phases, and made comments during group discussion. There was one report of a participant who experienced a brief period of nausea. 50 participants carried out reflective writing, and 57 answered the MEQ30 afterward. For the 7 participants who did not answer the MEQ30, one was an expert who we judged to be too familiar with the methodology. At least two others

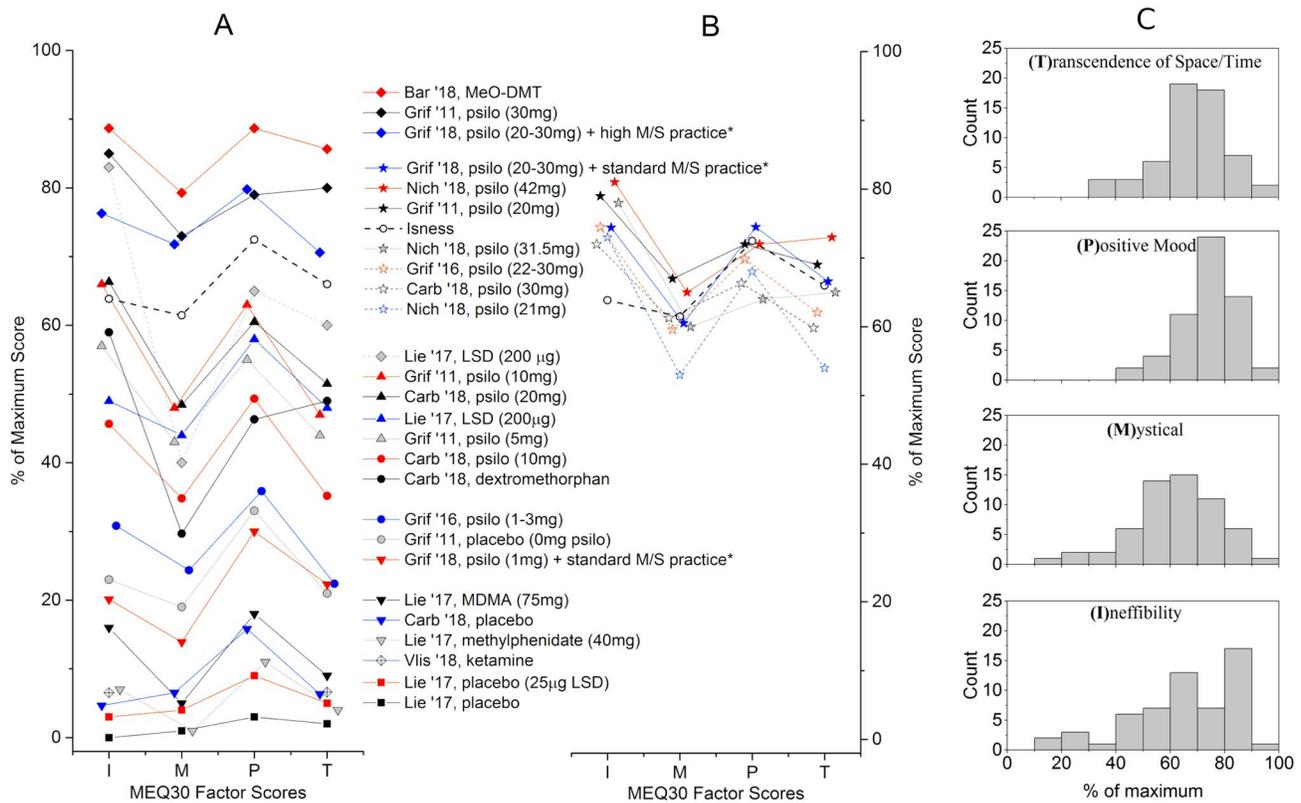

**Figure 2:** Comparison of the avg Isness factor scores to previously published ΨD research studies where the MEQ30 has been utilized (Bar '18 [69], Griff '11 [26], Griff '18 [27], Nich '18 [70], Griff '16 [28], Carb '18 [71], Lie '17 [72], and Vlis '18 [74]). The SM contains further details on the design of these studies. Panel (A) shows studies with at least 2 **M**, **P**, and **T** scores statistically distinguishable from Isness; (B) shows studies which are statistically indistinguishable from Isness; and (C) shows the distribution of the Isness MEQ30 results for each of the four factor scores. *M/S practice = 'Meditation & Spiritual' practice

indicated that they preferred not to answer quantitative questions in their post-Isness emotional state.

**Quantitative Analysis**
The MEQ30 (available in ref [1]) asks participants to rate the intensity with which they experienced 30 items on a 6-point scale [from "0 = none; not at all" to "5 = extreme (more than ever before in my life and stronger than 4)"], with three questions to capture ineffability **I**, fifteen mystical **M** [capturing unitive experiences, noetic quality, and sacredness], six positive mood **P**, and six transcendence of time/space **T**. Participant responses within each factor (**I**, **M**, **P**, **T**) are then averaged, and reported as a percentage of the maximum score. Fig 2A-B compares the Isness MEQ30 factor scores to the 26 previous studies in the altered states database [70] where the MEQ30 has been used to analyse both ΨD and non-ΨD altered states. Fig 2A-B shows that the MEQ30 can distinguish dose dependent effects of ΨDs. Tabulated data for Fig 2A-B is available in Table SM2.

Fig 2C shows the distribution of the Isness MEQ30 scores for each factor with a histogram bin width of 5%. Compared to the normally distributed **M**, **P**, and **T** scores, **I** is much noisier (see further analysis in Table SM3), and lower than might be expected. This may result from the fact that, by the time participants were given the MEQ30, they had undertaken 10 – 15 mins group discussion and reflective writing. So participants unable to find words in the immediate wake of their Isness 'peak experience' (i.e., a high **I** score) had ample opportunities to articulate their experience by the time they were given the MEQ30. Given these issues with the **I** scores we have restricted statements of statistical significance to the **M**, **P**, and **T** results.

In the absence of an experimental control group, we undertook comparative analysis of Isness to the previously published psychedelic studies in Fig 2A-B using independent sample t-tests with α = 0.05, following the approach Barsuglia et al. [3] used to analyze MEQ30 results obtained during uncontrolled MeO-DMT field tests. Fig 2 and Table SM2 show results of 26 different independent-sample t-tests, comparing the Isness MEQ30 results to each of the studies in Fig 2A-B. Despite its simplicity, the independent sample t-test gives results that are broadly aligned with more sophisticated statistical analyses described in Table SM3 & SM4. To make Fig 2, we classified a study as statistically indistinguishable from Isness if it at least two of its **M**, **P**, or **T** factor scores were statistically indistinguishable from the corresponding Isness factor scores. Compared to Isness, Fig 2 depicts:

- *3 published studies which are more intense*. These include: (1) a MeO-DMT study; (2) a 30 mg psilocybin study; and (3) a 2018 study by Griffiths et al [26] where participants in a program offering high levels of support

| Theme (N statements) | Indicative Quotations |
|---|---|
| **Positive Emotions (42)** | - [I] cried quite a bit, beginning [when] we all turned red… peaking when the lights went out and we laid down… beautiful<br>- [Isness is] calming, and it's relaxing, and it's transfixing<br>- [Isness] took me away from all of my horrible issues that I'm having at the moment. |
| **Connectedness (40)** | - A beautiful way to connect… not really connecting with anyone's stories or anyone's appearance… just the simple pure presence of another being… very touching.<br>- [Isness] made me think about our connection with all living beings including plants and animals. quite a paradigm shift |
| **Ego-dissolution (35)** | - Not knowing who the person [was] next to me allowed me to connect without any pre-judgment… I no longer saw others through the lens of beauty, social structure, age, etc… Rather, it was a pure joy of just being and connecting through light.<br>- You become all anonymous, just beings of light. It allows you to let go of your own sense of ego |
| **Embodied perceptual awareness (35)** | - One of the ways you become less self-conscious is to be conscious of your senses. this makes you really conscious of your senses.<br>- Felt so light [weightless] during the experience, and heavy afterwards<br>- I feel fully in my body & waves of energy coming over me. I feel the light in my body & I feel present with the other light beings in my environment. |
| **Reflection on Death (28)** | - The end felt like a peaceful death. The darkness & stillness at the end felt so peaceful<br>- Thought [about] my own death… the energy you give out to the world will always be there<br>- Felt connected to my dad who died 4 years ago |
| **Supportive Setting (25)** | - I've been very gently guided through a journey of light, stillness and silence<br>- Found the preparation very effective. Its ceremonial quality helped tune into the experience. Super important. |
| **Noetic Quality (20)** | - Felt so real. More real than real at times… a felt experience of the [knowledge] I already had of the light inside people and beings<br>- First sensation was a huge familiarity and I felt like crying…Peace… Feeling of coming home.<br>- really did make my heart burn with appreciation for the simple fact that I exist |
| **Transcendence of Space & Time (20)** | - Time runs differently in there.<br>- The vastness of space, the non-linear time, care and gentleness to other humans in this experience, and the self-love, helped push away the torture of existential anxiety |
| **Insights for everyday life (19)** | - You can radiate light out. And I liked the fact that it was the pressure that made it radiate. Sometimes when things are hard you need to add more pressure to be more light.<br>- The light of the mudra will stay with me for a long while. |
| **Sense of Beauty (18)** | - Beautiful, gentle and heart opening.<br>- A wonderful and magical experience. |
| **Comparison to Other Altered States (15)** | - Similar to a 5-MEO-DMT experience… pure energy, pure being, with no reference to physical space<br>- Transporting and illuminating… like the quiet point of an acid trip where calm beauty settles in<br>- Reminds me of mescaline… like a pandora's box has been opened and the universe revealed in its true form, rather than the virtual model created in our own heads. You're seeing a deeper form & fundamental truth |
| **Ineffability (9)** | - Ok. Words. Huh. I know some words. Huh.<br>- [I] feel moved, touched, quite quiet, words are hard to muster – I appreciate that because it indicates that I've been brought into my body, into my heart/energetic being. |
| **Childlike enchantment (7)** | - [I had a] Beautiful joyous innocent but also intimate feeling I don't remember feeling since I was a child.<br>- We were like children again, exploring and no judgement or anything. |
| **Metaphors for everyday life (4)** | - Playing with the [energetic thread] seemed like we were in control, but when [we stopped], the [thread] did its own stuff. It was an acceptance of the co-creation of environment and selves. It was astonishingly calm when we relaxed at the end… a powerful step down of need to control. |

**Figure 3: themes which emerged from inductive analysis as described in the text, along with indicative quotations. The themes are ranked according to the number of participant statements, N, which could be assigned to each theme**

in carrying out meditation and spiritual (M/S) practice were given 20-30 mg psilocybin.

- *7 published studies which are indistinguishable*. All of these studies involved participants being administered moderate to large doses (20-42 mg) of psilocybin. In one of these studies, Griffiths et al. [26] gave 20-30 mg psilocybin to participants involved in a program offering M/S practice support.
- *16 published studies which are less intense*. Amongst these, 6 were baseline studies, and 10 were drug-administration studies. The baseline studies included: 4 placebo studies; a 2016 study where Griffiths *et al.* gave participants sub-perceptual (1-3 mg) doses of psilocybin [25]; and a 2018 study where Griffiths *et al* [26] gave 1 mg psilocybin to participants in a program offering M/S practice support. The ten drug-administration studies investigated: MDMA [49, 69]; methylphenidate [49, 69]; ketamine [77]; dextromethorphan [6]; psilocybin [6, 24], and LSD [49].

Griffiths and co-workers identify an MEQ30 respondent as having had a 'complete MTE' when each of the **I**, **M**, **P**, **T** factor scores are ≥60% of the maximum. [40] In general, the fraction of participants reporting a complete MTE is proportional to the ΨD dose. Barrett and Griffiths [2] reported a meta-analysis of high dose (30 mg/70 kg) psilocybin studies in 119 healthy volunteers [22, 24, 26], and observed that **57%** of participants had 'complete' MTEs. Recent studies on 5-MeO-DMT (Fig 2C) [3] reported **75%** of MEQ30 participants having a complete MTE. Analysis of the Isness MEQ30 data indicates that **44%** of participants qualified as having a complete MTE. For comparison, Table SM1 shows complete MTE rates for those studies where it was reported.

### Qualitative Analysis

To better understand the quantitative data, we carried out qualitative analysis of participants': (1) group discussions after exiting VR, and (2) reflective writing. The group discussions were intended to enable maximum conversation amongst the participants, with the guide's role that of an active listener. The format of these group discussions was not rigidly prescribed; instead the emphasis was on the guide enabling conversation amongst the group. In many groups, participants did not immediately wish to speak. In such cases, the guide asked a simple question to initiate the conversation – e.g., 'How are you?'; 'How did that feel?'; 'What remains?'; 'How did you feel when the light faded?' Qualitative data analysis involved transcribing into digital format ~150 mins of recorded conversations (~14,000 words) and ~3,600 words of reflective handwriting. We then analysed this data by carrying out inductive thematic analysis, as outlined by Braun and Clarke. [5] The 14 themes outlined in Fig 3 were broadly similar across both the interviews and reflective writing, and enabled us to classify ~95% of the transcribed text. The themes in Fig 3 are ranked in order of how many participant statements could be attributed to each, along with some indicative quotations which correspond to each theme. The entire classification is available in the SM.

## 6. DISCUSSION

In devising Isness, we imagined VR as a form of psychedelic technology, investigating whether it could be used as a tool for eliciting subjective accounts of MTEs. To guide our design process for constructing the multi-person VR phenomenology, we looked to concepts which characterize MTEs. Quantitative analysis shows that, given a supportive setting, Isness produced 'peak experience' MTEs which are statistically indistinguishable from the MTEs observed in previous studies administering moderate to high doses of classical ΨDs. The qualitative analysis helps to rationalize the quantitative MEQ30 scores, providing insight into the subjective experiences and themes which arose. The group aspect of Isness appears to have amplified participants' sense of ego-dissolution, releasing them from the projections associated with typical social interactions, enabling them to '*connect without any pre-judgment… no longer [seeing] others through the lens of beauty, social structure, age, etc*'. The resultant state was '*like [we were] children again, exploring [without] judgement or anything*'. This encouraged a form of '*pure presence*', making participants '*really conscious of their senses*' and giving them space to '*just be*'. By focusing awareness on '*the light in my body and [feeling] present with the other light beings*', Isness left participants calm, relaxed, '*moved, touched, and quiet*'.

The dissolution of ego is the loss of what makes our personality distinct, and represents a kind of experience of death, [41, 76] likely explaining participants' comments that '*the end felt like a really peaceful death*', and observations that '*It's just a physical body, the energy will remain [after dying], and it will stay in a connection even though this [body] is gone*'. Comments like these are aligned with insights from Timmerman et al., [76] who identified a correlation between near-death experiences and the sense of ego dissolution which arise for participants in DMT experiences. Broadly speaking, digital technologies struggle to deal with the existential phenomenology of death; [43, 44, 50, 51] instead focusing on digital heirloom logistics (e.g., managing social media accounts for the deceased). The ego-dissolution achieved in Isness is an interesting contrast to social media, which enables connection, but encourages individuals to amplify their individual identity, often producing anxiety.

Comments like '*I've been very gently guided through a journey of light, stillness and silence*' draw attention to the significance of set and setting for thinking about how participants in VR experiences attribute meaning to an experience like Isness. Clearly, subjective attributions of meaning to Isness (and within HCI more broadly) depend on the supportiveness of the context framing the experience, and the broader conceptual framework in which it is embedded. The importance of set and setting is aligned with the idea that VR participants enter into a sort of 'hyper-suggestable' state.

The features which characterize MTEs provide a rich cross-disciplinary conceptual framework for thinking about meaning in HCI. The empirical study presented herein shows how these concepts can be practically implemented so as to design the kinds of meaning-making experiences that HCI workers like Mekler and Hornbaek, [56] Light et al, [50, 51] and Kaptelinin [43, 44] have outlined theoretically. Fig 3 and the SM provide a glimpse of the meaning and insight which participants attributed to their experiences. They reported '*a felt sense of [the] knowledge [I] already had*', and felt they saw '*the universe revealed in its true form… this absolute truth that we are made of light*'. Several participants cited insights and practices which they intended to carry into the world 'out-there', stating for example '*I will be using mudras more in my daily life*', expressing their intentions to '*remember this [experience], and in difficult situations connect back to [the] feeling of being able to create and connect*' with the luminosity of the material world, and observing how '*the pressure made [the mudras] radiate… Sometimes when things are hard you need to add more pressure to be more light*'. Several participants said they cried; one remarked '*I hate meditating but you make me want to try.*'

Isness differs from psychedelic psychotherapy studies in some important ways. For example, Isness preparation lasted ~15 mins for a group of four, far less than the preparation for studies carried out by Griffiths et al (see Table SM2) which typically include a total of 4 – 8 individual sessions (occurring before and after drug administration). Additionally, Isness's three phases last a total of ~70 minutes, considerably shorter than typical psilocybin and LSD experiences, which last anywhere from 6 – 14 hours. This may account for participants' comparisons to DMT, which usually lasts less than 30 mins. Finally, Isness was constructed as a group experience, whereas most ΨD studies are individual experiences.

Compared to other ΨD studies in Fig 2 (and Table SM1 and SM2), the N=57 Isness sample size was relatively large; however, this study had a number of limitations. For example, given that we did not carry out a control experiment, we are unable at this stage to make clear statements about the extent to which sample selection bias may have influenced our results. Our ability to make meaningful comparisons with previous studies depends on the assumption that the baseline MEQ30 responses of our participant sample are not anomalously high, and within the range spanned by 6 previously published baseline studies. Fig 2A shows that these baseline studies have a broad MEQ30 score distribution; however, our comparative statistical analyses (Fig 2B and Table S2) show that the Isness results are more intense ($p < 1E-12$) than all of them, including for example the 2018 study where Griffiths *et al* [26] gave 1 mg psilocybin to participants who were engaged in a supportive program of meditation and spiritual practice. At this stage, it is unclear how exactly to define a placebo for an experience like Isness, but this is an issue we will investigate in future work. We note that participants may have been more willing to provide mystical-type responses during group discussion, reflective writing, and during the MEQ30 analysis because of their belief (whether accurate or not) that we wanted such responses. On the other hand, because participants were drawn from a psychedelic conference, they were able to compare Isness to previous ΨDEs. For example, some participants specifically indicated that they were unable to rate the MTE occasioned by Isness as highly as previous ΨDEs. We also cannot rule out that the results reported herein are subject to the so-called 'winner's curse' [68] where effect sizes observed in trials of new treatments are inflated due to a variety of subtle effects. The design of Isness as a group experience suggests that the individual data may be correlated. In future studies we wish to investigate the correlation of the results obtained for participants within specific groups, and compare *intra*-group results to *inter*-group results.

Interesting questions arise concerning the 'authenticity' of a VR-occasioned MTE like that which arose during Isness. The term 'virtual reality' *per se* almost trivializes any MTE that emerges. Interestingly, the question of 'authenticity' has similarly been raised with respect to ΨDEs – i.e., there is a critique that ΨDs represent a sort of *virtual* form of MTE, lacking 'authenticity' compared to those which arise without drugs. [45] In a recent paper designed to evaluate this question through analysis of questionnaires from 2000 participants, Griffiths *et al* [27] showed that whilst mystical 'God encounter' experiences occasioned with and without ΨDs have some differences, the descriptive details, interpretation, and consequences of these experiences are markedly similar. Because MTEs occasioned by ΨDs are so similar to non-drug MTEs, Griffiths *et al* argue that it is problematic to assert that one is virtual and the other is not. This logic, combined with our evidence that VR can occasion MTEs, suggests that 'virtual reality' may be a concept best understood from a wider vantage point, where head mounted displays (HMDs) simply represent one kind of 'virtual reality' technology amongst a broader continuum of VR technologies, which include for example ΨDs, mythologies, rituals, meditation practices, lucid dreaming, [46] etc. By comparing the MTEs arising with VR and ΨD 'technologies', we can begin to understand where each sits within the broader spectrum of psychedelic technologies and VR technologies.

The approach outlined herein, where aesthetic hyperparameters defining a particular journey can be precisely defined, offers an interesting complement to ongoing ΨD research, opening up a range of further research directions. [58] It will be interesting to explore how Isness variants, constructed by exploring different domains of the aesthetic hyperparameter space, compare to different ΨDEs. Enabling new phenomenological experiences by designing new ΨD molecules is difficult, and clinical research efforts therefore focus primarily on tuning set and setting. Isness enables direct tuning of aspects of the participants' phenomenological experience, and different classes of experience almost certainly exist within the hyperparameter space. This idea has some analogy with proposals by Carhart-Harris et al., [8] who suggested that consciousness can access an ensemble of different metastable brain states, each of which uniquely influences aspects of perception. In future work, we hope to analyze Isness peak experience using a wider range of psychometrics (beyond MEQ30), and also to carry out longer-term follow-ups to better understand its impact. We also hope to understand how the neurophysiological impacts of Isness relate to EEG and fMRI observations during ΨDEs.

## 7. CONCLUSIONS

Within a supportive setting and conceptual framework, we have presented evidence suggesting that it is possible to design phenomenological experiences using multi-person VR which create the conditions for MTEs from which participants derive insight and meaning. Given that colloquial usage of the term 'psychedelic' is linked to drugs, we have imagined different words for describing a technology like Isness. Inspired by the Latin *numen* ('arousing spiritual or religious emotion; mysterious or awe-inspiring') and the Greek *pneuma* ('breath', 'spirit', or 'soul'), perhaps technologies like Isness may be described as *numedelic* ('spirit-manifesting', or 'spirit-revealing'). Analogous to psychedelic psychotherapy, we may imagine Isness as a tool enabling *numedelic psychotherapy*. Much ΨD research aims to help patients deal with addictions and end-of-life anxiety, individual conditions which represent the broader problems facing our culture right now. In a supportive therapeutic context, *numedelic* technologies like Isness may offer an opportunity for a digital culture which is addicted to unhealthy economic growth narratives to meditate on its own mortality.

## ACKNOWLEDGMENTS

DRG is supported by the Leverhulme Trust (Philip Leverhulme Prize), Royal Society (URF/R/180033) and EPSRC (institutional sponsorship award & EP/P021123/1); OM by EPSRC (EP/N00616X/2); MW by the Royal Society (RGF/EA/181075) and BBSRC (BB/R00661X/1); and JEP, MW, & TdH by the ArtSci International Foundation. People who provided support and useful conversations at various stages include: David Luke, Aimee Tollan, Andy Milns, Jae Levy, Chris Timmerman, Michael B. O'Connor, Alexander Jamieson-Binnie, Jonathan Barnoud, Helen Deeks, Alex Jones, Becca Walters, Rhosyln Roebuck-Williams, Joseph Hyde, Thomas Mitchell, Lisa Thomas, the Arts Council England Figuring team, Hester Jones, Michael Hoezl, & the Breaking Convention attendees who participated.

**SUPPLEMENTAL MATERIAL (SM)**

This SM includes:
(1) **Table SM1**, which outlines the experimental conditions and results of previously published ΨD studies
(2) **Table SM2**, which outlines the results of the independent sample t-tests which we carried out comparing Isness to each of the 26 previously published ΨD studies outlined in Fig 2A-B, and SI Table 1
(3) **Table SM3**, the results from normality tests carried out analyzing the distribution of the Isness factor scores
(4) **Table SM4**, the results obtained from hypothesis testing comparing Isness with previously published MEQ30 studies using methods which are more sophisticated than the independent sample t-test results in Table SM2, and which do not require the assumption of a normal distribution
(5) **Table SM5**, the raw MEQ30 I, M, P, and T scores obtained from Isness participants
(6) The full list of statements (obtained from transcribed group discussions and reflective writing) which were assigned to each of the respective themes in **Figure 3** of the main text. Statements from group discussions are in normal text. *Statements from reflective writing are in italics*.

| Author | Setting/context | Participants | Substances Tested (dose) | % complete ME |
|---|---|---|---|---|
| Nicholas 2018 (Nich '18) | Aesthetic living-room environment (ALRE); 4 sessions with guides before; integration meeting afterward | Healthy | 1. psilocybin (21 mg/70 kg) 2. psilocybin (31.5mg/70kg) 3. psilocybin (42 mg/70 kg) | N/A |
| Liechti 2017 (Lie '17) | Standard hospital patient room, with regular blood samples taken | Healthy | 1. Placebo (gelatin capsule) 2. LSD (200 μg) | 1. N/A 2. 12.5% |
| | Standard hospital patient room, with i.v. for regular blood sampling | Healthy | 1. MDMA (75 mg) 2. Methylphenidate (MPD) (40 mg) | N/A |
| | Embedded in 6–8 psychotherapy sessions | Terminal illness | 1. Placebo LSD dose (25 μg) 2. LSD (200 μg) | 1. N/A 2. 17% |
| Carbonaro 2018 (Carb '18) | ALRE; meet session monitor twice beforehand & twice afterward | Healthy | 1. Placebo (lactose or crystalline cellulose) 2. Dextromethorphan (400 mg/70kg) 3. psilocybin (10 mg/70 kg) 4. psilocybin (20 mg/70 kg) 5. psilocybin (30 mg/70 kg) | 1. 0% 2. 0% 3. 0% 4. 20% 5. 40% |
| Griffiths 2018 (Grif '18) | Two drug sessions, embedded in a 6-8 month support program combining meditation & spiritual practice | Healthy | 1. Low Dose psilocybin (1 mg/70 kg) + Standard M/S practice support 2. High Dose psilocybin (20-30 mg/70 kg) + Standard M/S practice support 3. High Dose psilocybin (20-30 mg/70 kg) + High M/S practice support | 1. 0%, 4% 2. 48%, 50% 3. 44%, 52% |
| Barsuglia 2018 (Bar '18) | psychospiritual retreat | Healthy | MeO-DMT (5 – 7 mg) | 75% |
| Griffiths 2011 (Grif '11) | ALRE; four 2-hour sessions with guides beforehand; one 2-hour session afterward | Healthy | 1. Placebo (0 mg/70 kg psilocybin) 2. psilocybin (5 mg/70 kg) 3. psilocybin (10 mg/70 kg) 4. psilocybin (20 mg/70 kg) 5. psilocybin (30 mg/70 kg) | 1. 6% 2. 11% 3. 17% 4. 61% 5. 67% |
| Griffiths 2016 (Grif '16) | ALRE; embedded in ~8 meetings with supportive session monitors | End-of-life cancer diagnoses | 1. psilocybin (1 – 3 mg/70 kg) 2. psilocybin (22 – 30 mg/70 kg) | N/A |
| Vlisides 2018 (Vlis '18) | Hospital Room, connected to i.v. and monitored using EEG | Healthy | Ketamine (0.5 mg/kg via i.v. over 40 min) | N/A |

**Table SM1:** details outlining the study protocols and complete mystical experience results for the 26 previously published ΨD studies outlined in Fig 2A-B, and Table SM2

|  | n | Ineffability* | | | Mystical | | | Positive Emotions | | | Transcendence of Space & Time | | |
|---|---|---|---|---|---|---|---|---|---|---|---|---|---|
|  |  | mean | std dev | p value | mean | std dev | p value | mean | std dev | p value | mean | std dev | p value |
| Isness | 57 | 63.9 | 20.4 | n/a | 61.5 | 16.5 | n/a | 72.5 | 11.4 | n/a | 66.0 | 13.4 | n/a |
| Bar '18, MeO-DMT | 20 | 88.7 | 12.6 | 1.3E-06 | 79.3 | 18.4 | 0.00007 | 88.7 | 11.7 | 3.4E-07 | 85.7 | 13.8 | 1.6E-07 |
| Grif '11, psilo (30mg) | 18 | 85.0 | 25.0 | 0.00026 | 73.0 | 25.0 | 0.01318 | 79.0 | 21.0 | 0.04736 | 80.0 | 25.0 | 0.00146 |
| Grif '18, psilo (20-30 mg) + high M/S practice | 25 | 76.3 | 19.5 | 0.00591 | 71.8 | 18.0 | 0.00650 | 79.8 | 19.0 | 0.01691 | 70.6 | 18.5 | **0.10320** |
| Grif '11, psilo (20mg) | 18 | 79.0 | 25.0 | 0.00566 | 67.0 | 21.0 | **0.12501** | 72.0 | 25.0 | **0.45454** | 69.0 | 30.0 | **0.27552** |
| Nich '18, psilo (42mg) | 12 | 81.0 | 26.0 | 0.00706 | 65.0 | 35.0 | **0.29655** | 72.0 | 28.0 | **0.46065** | 73.0 | 31.0 | **0.10615** |
| Carb '18, psilo (30mg) | 20 | 72.0 | 18.8 | 0.06083 | 61.3 | 21.0 | **0.48258** | 66.3 | 20.0 | 0.04847 | 59.8 | 17.0 | **0.05220** |
| Grif '18, psilo (20-30 mg) + standard M/S practice | 25 | 74.4 | 31.5 | 0.03695 | 60.5 | 36.5 | **0.43395** | 74.5 | 21.5 | **0.29056** | 66.6 | 28.0 | **0.44621** |
| Nich '18, psilo (31.5mg) | 11 | 78.0 | 26.0 | 0.02413 | 60.0 | 35.0 | **0.41360** | 64.0 | 29.0 | 0.04977 | 65.0 | 31.0 | **0.43172** |
| Grif '16, psilo (22-30mg) | 50 | 74.5 | 26.0 | 0.00998 | 59.6 | 29.8 | **0.34037** | 69.8 | 27.0 | **0.24917** | 62.1 | 23.9 | **0.14629** |
| Nich '18, psilo (21mg) | 10 | 73.0 | 27.2 | 0.10923 | 53.0 | 37.0 | **0.11708** | 68.0 | 28.0 | **0.19088** | 54.0 | 33.0 | 0.02488 |
| Carb '18, psilo (20mg) | 20 | 66.3 | 23.3 | 0.32742 | 48.5 | 28.0 | 0.00735 | 60.5 | 24.0 | 0.00207 | 51.5 | 21.0 | 0.00033 |
| Grif '11, psilo (10mg) | 18 | 66.0 | 25.0 | 0.35723 | 48.0 | 25.0 | 0.00494 | 63.0 | 25.0 | 0.01408 | 47.0 | 25.0 | 0.00004 |
| Lie '17, LSD (200μg) | 11 | 49.0 | 26.5 | 0.01952 | 44.0 | 26.5 | 0.00090 | 58.0 | 26.5 | 0.00198 | 48.0 | 23.2 | 0.00034 |
| Grif '11, psilo (5mg) | 18 | 57.0 | 30.0 | 0.13670 | 43.0 | 21.0 | 0.00012 | 55.0 | 30.0 | 0.00022 | 44.0 | 30.0 | 0.00002 |
| Lie '17, LSD (200μg)-A | 16 | 83.0 | 12.0 | 0.00032 | 40.0 | 24.0 | 0.00025 | 65.0 | 16.0 | 0.01905 | 60.0 | 20.0 | **0.08208** |
| Carb '18, psilo (10mg) | 20 | 45.7 | 19.4 | 0.00042 | 34.8 | 19.8 | 4.7E-08 | 49.3 | 18.0 | 2.1E-09 | 35.2 | 16.2 | 1.2E-12 |
| Carb '18, dextromethorphan | 20 | 59.0 | 18.0 | 0.17416 | 29.7 | 21.0 | 7.1E-10 | 46.3 | 22.1 | 1.3E-09 | 49.0 | 21.2 | 0.00004 |
| Grif '16, psilo (1-3mg) | 50 | 30.8 | 31.7 | 1.5E-09 | 24.3 | 27.1 | 2.7E-14 | 35.8 | 28.3 | 5.6E-15 | 22.4 | 20.5 | 2.8E-24 |
| Grif '11, Placebo (0mg psilo) | 18 | 23.0 | 21.0 | 1.1E-10 | 19.0 | 21.0 | 1.4E-13 | 33.0 | 21.0 | 4.2E-16 | 21.0 | 30.0 | 1.3E-13 |
| Grif '18, psilo (1 mg) + standard M/S practice | 25 | 20.1 | 20.5 | 6.0E-14 | 13.9 | 17.5 | 1.6E-19 | 30.0 | 18.0 | 1.5E-21 | 22.3 | 20.5 | 7.0E-19 |
| Carb '18, Placebo | 20 | 4.7 | 9.0 | 2.4E-20 | 6.5 | 9.3 | 4.2E-23 | 15.8 | 11.7 | 1.2E-30 | 6.3 | 9.8 | 1.3E-29 |
| Lie '17, MDMA (75mg) | 30 | 16.0 | 16.4 | 1.7E-18 | 5.0 | 32.8 | 9.4E-18 | 18.0 | 21.9 | 2.0E-26 | 9.0 | 27.3 | 2.2E-22 |
| Lie '17, Placebo (25μg LSD) | 4 | 3.0 | 0.0 | 8.7E-08 | 4.0 | 0.0 | 1.8E-09 | 9.0 | 8.0 | 4.6E-16 | 5.0 | 2.0 | 5.2E-13 |
| Lie '17, methylphenidate (40mg) | 30 | 7.0 | 16.4 | 1.5E-22 | 1.0 | 0.0 | 2.9E-34 | 11.0 | 21.9 | 8.0E-30 | 4.0 | 0.0 | 1.4E-41 |
| Lie '17, Placebo | 16 | 0.0 | 0.0 | 7.5E-20 | 1.0 | 0.0 | 2.2E-23 | 3.0 | 0.0 | 2.3E-36 | 2.0 | 0.0 | 7.5E-30 |
| Vlis '18, ketamine | 15 | 6.5 | 2.6 | 7.1E-17 | n/a | n/a | n/a | n/a | n/a | n/a | 6.6 | 2.1 | 7.1E-27 |

**Table SM2: p-values from independent sample t-tests results comparing the Isness MEQ30 I, M, P, and T factor scores to the 26 previously published ΨD studies given in Fig 2A-B and Table S1, along with the data used to carry out these tests. Fields in bold text indicate factor scores which are statistically indistinguishable (p > 0.05) from the corresponding Isness factor score. Studies highlighted in grey are those for which at least two of the M, P, or T scores are statistically indistinguishable from Isness. Studies highlighted in blue are previously published baseline studies, which are outlined in detail in the text. *The I factor score results are included here for the sake of completeness, but we did not include a discussion of their statistical significance in the main text, given their noisy distribution.**

**Testing the normality of the Isness I, M, P, and T factor score distributions.** Table SM3 outlines tests which we performed on the Isness I, M, P, and T factor scores to evaluate whether the data was characteristic of a normal distribution. These tests were carried out in Mathematica using the command 'DistributionFitTest' with a statistical significance level of $p > 0.05$. This command runs all appropriate statistical tests for normality and returns a $p$ value and conclusion.

| Test | Ineffability | Mystical | Positive Emotions | Transcendence of Space & Time |
|---|---|---|---|---|
| Anderson-Darling | 0.00071 | **0.11383** | **0.11338** | 0.02928 |
| Baringhaus-Henze | 0.00680 | **0.10440** | **0.20458** | **0.06777** |
| Cramer-von Mises | 0.00134 | **0.23670** | **0.08265** | 0.04626 |
| Jarque-Bera ALM | 0.04780 | **0.07582** | **0.28772** | **0.11715** |
| Mardia Combined | 0.04780 | **0.07582** | **0.28772** | **0.11715** |
| Mardia Kurtosis | **0.91112** | **0.65614** | **0.52726** | **0.40964** |
| Mardia Skewness | 0.01230 | 0.03858 | **0.31322** | **0.13949** |
| Pearson chi-squared | 0.00325 | **0.63544** | 0.00778 | **0.19484** |
| Shapiro-Wilk | 0.00194 | **0.05635** | **0.31441** | 0.04670 |

**Table SM3:** Results ($p$ values) from various normality tests for the datasets. Where the test would not reject the null hypothesis (that the data is a normal distribution) at the p>0.05 level the p values are in bold. The results clearly shown that the I dataset fails nearly all of the tests, whereas the M, P, and T scores pass most tests.

**Comparing the Isness I, M, P, and T factor scores to previously published studies.** Table SM4 shows the results of statistical hypothesis tests evaluating whether the Isness I, M, P, and T factor scores were statistically indistinguishable from previous MEQ30 studies. We carried these tests out using the Mathematica command 'LocationTest' on the data measured Isness factor score $\mu$ (ineffability (I), mystical (M), positive emotion (P), and transcendence of space and time (T)). We compared to $\mu$ to $\mu_0$, where $\mu_0$ was taken as the mean of the respective literature factor scores in Table SM2. For any given factor score, the null hypothesis was that $\mu = \mu_0$, i.e. the Isness factor score is statistically indistinguishable from a literature factor score. The alternative hypothesis was that $\mu \neq \mu_0$, i.e. that there was a statistically significant difference between the Isness factor score and the literature factor score. To carry out these tests, we utilized the complete set of Isness factor scores given in Table SM5, and allowed Mathematica to choose the most appropriate test. For the I, P and T data, Mathematica selected the Sign test, which compares the respective medians of two datasets. The Sign test does not assume the Isness data is normally distributed and is thus more appropriate for dealing with datasets that include outliers. Because we did not have the individual scores from previous studies (i.e., the equivalent to the data in Table SM5), we assumed that scores obtained in previous studies were normally distributed, with a median value close to the mean. Mathematica determined that the M dataset was sufficiently normal to use the t-test, which compares the respective means of each dataset.

| | n | Ineffability p value | Mystical p value | Positive Emotions p value | Transcendence of Space & Time p value |
|---|---|---|---|---|---|
| Isness | 57 | n/a | n/a | n/a | n/a |
| Bar '18, MeO-DMT | 20 | 2.3E-14 | 4.5E-11 | 4.3E-13 | 4.3E-13 |
| Grif '11, psilo (30mg) | 18 | 4.2E-09 | 2.3E-06 | 0.00126 | 7.1E-10 |
| Grif '18, psilo (20-30 mg) + high M/S practice | 25 | 0.01635 | 0.00002 | 0.00126 | **0.06274** |
| Grif '11, psilo (20mg) | 18 | 0.01635 | 0.01501 | **0.59664** | **0.59664** |
| Nich '18, psilo (42mg) | 12 | 4.2E-09 | **0.11682** | **0.59664** | **0.06274** |
| Carb '18, psilo (30mg) | 20 | **0.59664** | **0.91868** | 0.00001 | 0.00001 |
| Grif '18, psilo (20-30 mg) + standard M/S practice | 25 | 0.01635 | **0.64079** | **0.79137** | **1.0** |
| Nich '18, psilo (31.5mg) | 11 | 0.01635 | **0.48791** | 0.00001 | **0.89395** |
| Grif '16, psilo (22-30mg) | 50 | 0.01635 | **0.38179** | 0.00320 | 0.00320 |
| Nich '18, psilo (21mg) | 10 | **0.59664** | 0.00025 | 0.00320 | 2.7E-08 |
| Carb '18, psilo (20mg) | 20 | **0.11116** | 1.7E-07 | 7.5E-07 | 5.7E-10 |
| Grif '11, psilo (10mg) | 18 | **0.11116** | 7.2E-8 | 3.3E-06 | 5.7E-10 |
| Lie '17, LSD (200µg) | 11 | 0.00001 | 6.9E-11 | 5.7E-10 | 5.7E-10 |
| Grif '11, psilo (5mg) | 18 | 0.01635 | 1.2E-11 | 5.9E-12 | 5.7E-10 |
| Lie '17, LSD (200µg)-A | 16 | 4.2E-09 | 7.5E-14 | 0.00001 | 0.00013 |
| Carb '18, psilo (10mg) | 20 | 1.5E-7 | 1.8E-17 | 2.3E-14 | 4.3E-13 |
| Carb '18, dextromethorphan | 20 | 0.01635 | 1.0E-20 | 8.0E-16 | 5.7E-10 |
| Grif '16, psilo (1-3mg) | 50 | 6.4E-11 | 7.7E-24 | 1.4E-17 | 1.4E-17 |
| Grif '11, Placebo (0mg psilo) | 18 | 5.9E-12 | 1.3E-26 | 1.4E-17 | 1.4E-17 |
| Grif '18, psilo (1 mg) + standard M/S practice | 25 | 5.9E-12 | 4.6E-29 | 1.4E-17 | 1.4E-17 |
| Carb '18, Placebo | 20 | 1.4E-17 | 3.0E-32 | 1.4E-17 | 1.4E-17 |
| Lie '17, MDMA (75mg) | 30 | 1.4E-17 | 7.4E-33 | 1.4E-17 | 1.4E-17 |
| Lie '17, Placebo (25µg LSD) | 4 | 1.4E-17 | 3.0E-33 | 1.4E-17 | 1.4E-17 |
| Lie '17, methylphenidate (40mg) | 30 | 1.4E-17 | 2.1E-34 | 1.4E-17 | 1.4E-17 |
| Lie '17, Placebo | 16 | 1.4E-17 | 2.1E-34 | 1.4E-17 | 1.4E-17 |
| Vlis '18, ketamine | 15 | 1.4E-17 | n/a | n/a | 1.4E-17 |

Table SM4: p-values obtained from compares the Isness MEQ30 I, M, P, and T factor scores to the 26 previously published ΨD studies given in Fig 2A-B and Table SM1. Fields in bold text indicate factor scores which are statistically indistinguishable (p > 0.05) from the corresponding Isness factor score. Studies highlighted in grey are those for which at least two of the M, P, or T scores are statistically indistinguishable from Isness. Studies highlighted in blue are previously published baseline studies, which are outlined in detail in the text.

| Participant number | Transcendence of Space & Time | Positive Emotions | Ineffability | Mystical |
|---|---|---|---|---|
| 1 | 63.33 | 70.00 | 60.00 | 64.00 |
| 2 | 63.33 | 70.00 | 53.33 | 61.33 |
| 3 | 96.67 | 83.33 | 86.67 | 86.67 |
| 4 | 33.33 | 63.33 | 33.33 | 30.67 |
| 5 | 70.00 | 60.00 | 53.33 | 65.33 |
| 6 | 60.00 | 76.67 | 66.67 | 46.67 |
| 7 | 60.00 | 76.67 | 20.00 | 69.33 |
| 8 | 76.67 | 83.33 | 66.67 | 68.00 |
| 9 | 63.33 | 70.00 | 40.00 | 60.00 |
| 10 | 70.00 | 50.00 | 26.67 | 44.00 |
| 11 | 73.33 | 83.33 | 80.00 | 71.43 |
| 12 | 63.33 | 66.67 | 40.00 | 68.00 |
| 13 | 53.33 | 66.67 | 66.67 | 54.67 |
| 14 | 63.33 | 80.00 | 66.67 | 68.00 |
| 15 | 60.00 | 60.00 | 60.00 | 56.00 |
| 16 | 73.33 | 76.67 | 73.33 | 58.57 |
| 17 | 70.00 | 66.67 | 66.67 | 50.67 |
| 18 | 56.67 | 76.67 | 66.67 | 73.33 |
| 19 | 66.67 | 73.33 | 60.00 | 62.67 |
| 20 | 43.33 | 56.67 | 40.00 | 41.33 |
| 21 | 80.00 | 86.67 | 80.00 | 82.00 |
| 22 | 73.33 | 80.00 | 80.00 | 54.67 |
| 23 | 70.00 | 73.33 | 73.33 | 74.67 |
| 24 | 65.00 | 76.67 | 80.00 | 67.14 |
| 25 | 73.33 | 80.00 | 46.67 | 69.33 |
| 26 | 56.67 | 66.67 | 73.33 | 56.00 |
| 27 | 63.33 | 70.00 | 73.33 | 61.33 |
| 28 | 73.33 | 70.00 | 66.67 | 58.67 |
| 29 | 76.67 | 83.33 | 86.67 | 78.67 |
| 30 | 63.33 | 78.33 | 80.00 | 73.71 |
| 31 | 56.67 | 70.00 | 66.67 | 48.00 |
| 32 | 85.00 | 76.67 | 93.33 | 66.67 |
| 33 | 33.33 | 56.67 | 13.33 | 20.00 |
| 34 | 83.33 | 90.00 | 86.67 | 76.00 |
| 35 | 86.67 | 83.33 | 86.67 | 78.67 |
| 36 | 66.67 | 66.67 | 53.33 | 50.67 |
| 37 | 56.67 | 70.00 | 80.00 | 48.00 |
| 38 | 63.33 | 53.33 | 53.33 | 33.33 |
| 39 | 60.00 | 60.00 | 53.33 | 54.67 |
| 40 | 80.00 | 83.33 | 86.67 | 81.33 |
| 41 | 33.33 | 46.67 | 20.00 | 22.67 |
| 42 | 73.33 | 76.67 | 80.00 | 81.33 |
| 43 | 73.33 | 80.00 | 80.00 | 50.67 |
| 44 | 63.33 | 80.00 | 73.33 | 77.33 |
| 45 | 43.33 | 73.33 | 46.67 | 44.00 |
| 46 | 92.00 | 86.67 | 73.33 | 81.33 |
| 47 | 53.33 | 41.67 | 13.33 | 19.33 |
| 48 | 80.00 | 70.00 | 66.67 | 61.33 |
| 49 | 60.00 | 61.67 | 53.33 | 58.67 |
| 50 | 43.33 | 60.00 | 46.67 | 52.00 |
| 51 | 66.67 | 73.33 | 53.33 | 68.00 |
| 52 | 80.00 | 100.00 | 100.00 | 81.33 |
| 53 | 63.33 | 70.00 | 73.33 | 58.67 |
| 54 | 73.33 | 76.67 | 80.00 | 74.67 |
| 55 | 73.33 | 76.67 | 80.00 | 72.00 |
| 56 | 76.67 | 76.67 | 80.00 | 74.67 |
| 57 | 70.00 | 96.67 | 80.00 | 94.67 |

**Table SM5:** Raw factor scores obtained from the Isness MEQ30 questionnaires

**Theme: Positive Emotions**

1. It's very meditative.
2. I found it quite moving.
3. Sense of calm.
4. It was nice – I made a mudra at someone and waved at someone and they waved back.
5. Yeah I'd like to live there. [laughing] Like to have one of these would be so calming you could just go into this other world. It's very calming.
6. Well it's calming, and it's relaxing, and it's transfixing.
7. [participants spontaneously hug one another]
8. Yeah it certainly opens the heart. My heart feels really warm and really open.
9. It was very gentle, wasn't it. Very delicate.
10. Soft.
11. [feeling] Wonderful.
12. Calm.
13. So calm.
14. I feel quite emotional actually.
15. Yeah. quite touched.
16. Sense of calm. But also just… hard to put to words. Like very connected to my breath.
17. it made we want to laugh, I don't know why. I think that's just my joy isn't it by seeing these things. It probably just evokes some sense of excitement and laughter in me.
18. [Isness] for me took me away from all of my horrible issues that I'm having at the moment. So that I thought it was a real beauty and I just really wanted to experience it.
19. [the lights fading felt] Sad. That made me feel really sad because I was loving it so much.
20. It's such a strange and beautiful… very different to what we know in the real world. in this world.
21. Yeah very symbolic also. With the light, and stepping into the circle. And being together in the light. And you look outside and it's dark. It was a good feeling.
22. I felt like I was going with everyone else. It just felt like a flow.
23. it brought a tear to my eye.
24. Very calm.
25. Everything is still vibrating a little bit. I can still feel that. But at a much higher frequency.
26. *Feelings of happiness afterwards.*
27. *An uplifting, immersive experience exploring us as the light and energy we are.*
28. *I definitely cried.*
29. *All in all really enjoyed the experience.*
30. *A moving and desirable experience.*
31. *I felt a little scared before because I didn't know what to expect. As the experience continued I relaxed a lot more and felt barriers break down between me and the others. I guess that's what it's all about.*
32. *Not knowing where you are and losing control is not [maybe 'bit'] scary (like psychedelics)*
33. *Some mood swings going into the darkness and debts [sic – maybe they meant death?]*
34. *Delight at being able to make shapes and movement with the lines and waves and spheres provided. [maybe spaces, not spheres?]*
35. *I felt totally absorbed and then touching other peoples' light hands was such a happy feeling.*
36. *Being slow and gentle and tentative made me come out of it with a real sense of calm and big smile on my face.*
37. *The end, when the light goes off is interesting too, I was wondering if it would make me sad or scared, but it was peaceful and relaxing.*
38. *Feeling in general relaxed and was easy to surrender to it. After [the experience finished I was] feeling a little dizzy.*
39. *It felt light (i.e. not heave) and soft and warm*
40. *Sense input is calming. I was moved at times. Actually to tears.*
41. *Not gonna lie: I cried quite a bit, beginning with the moment we all turned red… peaking when the lights went out and we laid down… beautiful.*
42. *At first I was getting claustrophobic with the blindfold and the muggy air and I was worried about this but the headset didn't seem so much. More because of the wider sense of space I think.*

**Theme: Connectedness**

1.  It was lovely that you didn't know who's who. And it didn't matter. We're all the same.
2.  That's what I liked about it. Everybody is the same. Not having any judgements.
3.  That's a good place to start. It's very levelling. Everybody is equal in it. You're using these representations where you're all represented in the same way. That was really beautiful.
4.  There was definitely a few emotions in that felt presence of connectivity. Ummm… Yeah quite powerful.
5.  I loved the web of connection between us. And manipulating it and seeing how even if we're across the room from one another, my impact impacts all the way over there. And we have this like shared web that we're all populating with each other.
6.  And like figure out how we can all take our piece of the web, and together in a unified way move it together. Yeah.
7.  And just all the way through, to get to this idea of just being light as this core thing shared between everybody. That's nicely done. It works.
8.  So it just obviously shifted my worldview into this like we're all connected, and we all come from this one pure light source that we're all attached to and its feeding us and it's animating us
9.  It's a very nice perception that we're all absolutely the same and undifferentiable in some way and yet different in some other way. It's fun.
10. It felt very emotional in a good way when we juts came in and we saw everyone's lights come on. It was so beautiful. Like so gentle and loving to see everyone and their lights and all the oher lights, and just this feeling of connection.
11. And the music, it just felt so loving, and that feeling of connection and beyond the material, and that connection which I feel in my heart to be true, but it brings it back in that sense of the felt perception of it. Little cosmic strands.
12. Yeah I liked the connection between my effect and the effect of everything. Like kind of like I was dealing with how much I can affect, and how much I can create and how much is created by the others. And everything. And it was really calming. I feel very relaxed, yeah very meditative. I felt like it was easy to go into that space, to just get taken away with it. Yeah really enjoyed it.
13. I thought it was easier to connect cuz we're all the same. So I like the willingness there.
14. Incredible seeing the different energies move. Like yourself connecting with the others. Amazing.
15. I feel really calm as well. I think it was was really amazing to feel so connected. For me the connection really came out of it. It made me think of how everything's connected. And it was cool to see that visually and experience it.
16. No I could. I was smiling a lot. I just felt a connectedness to these, I know it's silly to make everything humanistic, but even the energy. I felt like the vibrations, I don't know
17. It's just a beautiful way to connect. And we're not really connecting with anyone's stories or anyone's appearance. It's just the simple pure presence of another being. And there's something very touching about that. It's very sweet.
18. I wanted to see if the light and the power came stronger together.
19. I think after I got over the fascination of "me", I wanted to draw everyone together. So I found myself sort of wanting to get all the – obviously I'm aware that it's hands that are creating the lights – I wanted to get all the hands to come together, and make something beautiful out of all the lights. Together. So that was quite emotional for me because it was like let's bring everything together.
20. I loved watching what I could do with the shapes, and now i had it before, but I have an even greater sense of love and empathy especially for you guys… But we just had a really strong connection there. When it went over my eyes, it was just like I love you, your beautiful, I love you, you're beautiful, I love you you're beautiful.
21. So with the index finger mudra you could control things, but with the middle finger you were a bit more passive and the world's happening around you. And then you stop moving and you undo the mudra and then you realize that everything just continues around you. And then you can join in again at any point and just become actively connected
22. I loved the web. In the beginning the web that was connecting everything. And on the ground.
23. *I felt joy at the feeling of unity achieved with total strangers, free of the cultural constraints that are usually present in most day-to-day interactions with other fellow human beings*
24. *Very connected to others in that we were one brain/being*
25. *I enjoyed the interactive aspect how we connect with each other as pure energy instead of solid state*
26. *It was the <u>cutest</u> way to connect with people sans judgment and the usual [might be 'visual'] things that get in the way of connection.*
27. *Loved the feeling of connection and that everyone was the same.*
28. *I'm left with a sense of connectedness.*
29. *I felt very connected with my fellow light beings and touched by the innocent and tender dance we shared.*
30. *It was really beautiful and it made me think about our connection with all living beings including plants and animals. It was quite a paradigm shift.*

31. *Connecting and beautiful. Feel connected to breath and remarkable to think of life and Self + others as light flowing through.*
32. *We were all the <u>same</u>.*
33. *The connection I felt to the other "beings" in the space was amazing and without the "veil" that is often over us in daily life*
34. *Sense of connection with other participants built very rapidly*
35. *Soft journey, wholesome – sense of oneness and wholeness. Both as an individual and as a group.*
36. *A real feeling of connectedness + compassion for the people in my group*
37. *I appreciated the subtlety and calm of the VR; the cords connecting us in the physical added an element of care for the others and made me want to hold each of them after we took off the headsets.*
38. *DEEP SENSE OF CALM AND CONNECTION TO NATURE.*
39. *I particularly enjoyed the possibility to connect with others via a very basic/fundamental concept, the theme of interconnectedness was prevalent throughout.*
40. *This was an exquisite experience of connection. I feel fully in my body & waves of energy coming over me. I feel the light in my body & I feel present with the other light beings in my environment. This technology feels nourishing, awakening, enlivening, and calming all at once. To manipulate the web of building blocks of creation, the primordial elements that animate life coming together, and seeing how interconnected all life is, and how every action we take impacts the space we share…Powerful.*

**Theme: Ego dissolution**

1. In general, lots of forgetting & trying to figure out who was who (eg Sun 15.00, Sat 14.00)
2. Yeah you had no feeling of self-consciousness really, because of the way you were represented anonymously. It's just freeing.
3. No social anxiety is a weird feeling. A nice break from the normal default.
4. You kind of had no tendency to show off or anything like that… Ego sort of dissolution. It's a collaborative thing. Each one trying to do the best for the group.
5. Yeah you want to group to have the best experience. And yeah if you can show someone something they're not quite getting, but there's no tendency to show off at all.
6. It's more a curiosity isn't it of seeing how it was and how it is.
7. Curiousity and playfulness for your own benefit, but then yeah trying to also create something beautiful to share with everyone else, but not in a grandiose I'm better way.
8. No. and there's no way to really attack the other people or make them have a negative experience, is there. There's no way you can annoy people in it. It's a very very positive collaboration. There's no competition element in there or anything like that. Or I want to get them all to myself.
9. Just seeing each other as pure full energetic beings and not knowing who was who in the space. I think I figured it out eventually. But I just not knowing that and being in that purity with each other in this expansive darkness.
10. I guess the preparation of stripping down everything. Like leaving whatever of the day behind. And this whole getting rid of your… even who you are. [identity]
11. And you didn't know who you were interacting with, there was no projections. You can't see, you don't know who you're interacting with.
12. [After Isness] The projections come back… There's more shyness. There's more a questioning of how you're perceived. Yeah. It's slightly less safe.
13. More assumptions. All your assumptions come back.
14. In there it's just light, no assumptions.
15. I think not knowing who you're touching in the space as well, you know, when you come back.
16. I think one of my favorite parts of the experience was when I could look down and not see myself, and just see the abyss. I really liked that, just looking up and seeing the heavens. So I quite liked the fact that I couldn't see my own light but I could see other people's. so I didn't feel the loss of my own light, because I couldn't see it anyway. I like that experience of just being nothingness and just being part of what I think was nature. But I think seeing everyone else's light and just seeing people as light and not as people I quite liked that and what the voice was saying about now imagine carrying that light around. I liked that. But I liked not being able to see myself and just be floating. That was really wonderful.
17. I didn't know who anybody was, just like all beings of light.
18. I loved we were all the same as well. The light heads and light hands. So pretty. Dancing together.
19. I'm very conscious of how people react to me, and how I react to them. And being freed from that self consciousness enabled me to act in a way that's much more playful than I would normally feel comfortable with.
20. I was happy that we were all the same.
21. For people who didn't do psychedelics... If people are scared, then have this experience, that could really help them to look at the world differently and let go of issues, boundaries, thoughts, beliefs.
22. Extraordinary sense of dissolution. Maybe. But in a very gentle way. Very nice feeling.
23. And just light. Not male, female, old, fat, young. There was nothing to make a judgement about. It was quite, quite liberating.
24. It's amazing also how you can have this contact, no? that is so intimate also with people that you don't know. That you've never met. And you interact. It's very primal. ..Yeah I don't think I could do that if I could see. So awkward.
25. And I like the way the lights anonymized us all.
26. You know we know each other, but then in this space you become all anonymous, just beings of light. It allows you to let go of your own sense of ego as well.
27. And I guess maybe in another scenario you would feel self conscious about flailing around. But I didn't feel any consciousness.
28. *A very special experience to feel and see each other and ourselves as only light. Brilliant and beautiful one and all – nothing to judge or feel as different. Very liberating, playful – each movement creating threads, the interconnected web.*
29. *Beautiful to experience being in flow with others – this is something I feel I would find it very difficult/impossible to get into ordinarily due to awkwardness) with the exception of if psychedelics were involved!)*
30. *Liberating – no sense of who others are – only that we are light. Felt power of Self – and freeing to just see as a light.*

31. *Sense of unselfconsciousness which enables playfulness without <u>social anxiety.</u> Liberation from normal social constructs/biases eg: race, sex etc.*
32. *Really enjoyed being disembodied and relieved of projections about others.*
33. *Not knowing who the person next to me has allowed me to connect without any pre-judgment. Could it be seen as clear connection? Removal of the body was immense. I no longer saw others through the lens of beauty, social structure, age, etc… Rather, it was a pure joy of just being and connecting through light. I hope that as many people can experience this moment… especially those who do not like each other.*
34. *The seeing others as pure energy was something that created an equality too – no judgments on any physical appearances or any preconceptions. Also removing your own anxiety and self consciousness that is all too often hindering connections with others in the material world.*
35. *The experience was strange. Very anonymising. Just light and energy. I was not concerned at all with who was who*

**Theme: Embodied Awareness**

1. The disembodiment is remarkable actually. Really.
2. It would have been nice not to have a body, because then there would be no bumping.
3. Even the carpet actually, feeling different textures under my feet.
4. Felt weightless… Especially because when we looked down, we didn't see the body, but you could see movement. So it felt like floating.
5. I noticed no one made a sound. No one laughed. No one said anything. It was all quiet.
6. But that's kind of cool. It's feels a bit like sense deprivation in a way. Because you are deprived of some senses. You're primarily visual, you're limited. But you can really focus on the small things that you still have. And you kind of realize that it's also a lot. And that is one way to switch away from others. It's interesting.
7. But maybe you are less yourself. You are more egoless in that space. Cuz you are, one of the ways in which you become less self conscious is to be conscious of your senses. So this makes you really conscious of your senses. Cuz you are disorientated you are much more embodied. If you know what I mean. So maybe it does shut down the default mode network in some ways. Because you are kind of just groping around in your senses. So you are more embodied.
8. I'm aware of myself as tall. So when we had to connect, what's hand level for me isn't hand level for everyone else. So that immediately starts to embody other people around me. So shoulder height for me is not shoulder height for everyone else. So it's a change of mindset when you make contact.
9. I found it easier to touch because I didn't know who I was touching
10. I was aware of that when I was playing a little bit. I was like hang on. I knew I was reaching somebody's hand. I was very careful. cuz you could see where their hands were, yeah, but then I suddenly thought well hang on is this ok? So I was aware well everyone's got different tolerances for contact. And I wanted contact. But it's good you know where the hands are.
11. You don't necessarily think of it as a hand. You just think of it as a light. It's just slightly different.
12. But I did wonder should I hold someone's hand? Or should I just make contact?
13. It would be good I think if everyone agrees there will be some contact. Then you know that everyone here is ok. I was slightly worried, wanting to play and explore, but wanting to not annoy or…
14. I think that the corporeality, what that did for me was – I tried moving my legs and I couldn't see them and that was great because I felt weightless. But as soon as we had to make contact I had a sense of scale again. So I think that's the effect. Cuz I know my own dimensions and my own weight and where things are in relation to me. But until then it's just floating, and you know it tracks the movement of your hands and head but not the rest of your body, and the scale was sometimes quite huge and sometimes quite small. But as soon as you have physical contact, I'm just aware of how things are again.
15. I agree, being aware of the thread actually is part of the experience….It made it more about that kind of interaction and being delicate. The delicateness of it is so kind of lovely.
16. Great when it went through the floor. That was good, it like's oh there's a whole.. You really felt like you were disembodied at that point. kind of oh right there's no floor either. I certainly can't see my feet. But you have a sense of where the floor is, and then that suddenly goes. Which is really nice.
17. The fact that I was with people I know specially menas there was freedom to go in and be more tactile. That was nice, we had some nice moments.
18. Yeah I feel like def a little dizzy now. But when I had it on, I didn't feel dizzy at all. More now, when it stopped.
19. I felt a bit disappointed when the light went out. The brightness was kind of calming in the darkness. So there was a lot sensation there. In the beauty of the light.
20. It was surprisingly real… [in the sense that] in the ordinary waking reality we know that it's not real. It's called virtual reality for a reason, right? But it feels… you can connect with the experience and treat it as you would any other experience. You can accept it like you can accept any other experience. And it feels real in that way.
21. At one point I thought I was surrounded by smoke [and started coughing]… when it was purple it was ok. But when it got grey, and then it was like I don't know, as if I was suffocating or I don't know what it was. But there was no real smoke. But that's I think because it was grey and really dense and all around me.
22. When I started coughing I noticed that yes I was separate from [the rendered smoke].
23. I actually smelled the smoke
24. You just felt like it was moving through me, which sounds a bit weird. But it was just amazing to not be in a body and to have things just floating. So yeah. Sounds a bit strange but.
25. And you feel a great sense of lightness, almost weightlessness… Right now I feel very heavy. And when I was shaping the volumes and space in front of me I wasn't aware a lot of my limbs.
26. It's [quite freeing]. And also to not see your own body is really amazing as well.

27. Like for me [as a man] I almost forgot that there were three women. You were just like light. I don't know, it was very cozy being close together.
28. And once we'd like made the initial joining [our mudras together] it felt like a progression to me. I was like why would I go back from that? Back to just having my two lights [on my own].
29. It's kind of amazing how little time it takes for your brain to accept what it's seeing there as reality.
30. *Felt so light during the experience, and heavy afterwards.*
31. *I feel very peaceful in my body, empty of thoughts.*
32. *The cables are a bit disturbing but the gloves and the mask were very comfortable.*
33. *Sometimes I felt constrained by the cables and gloves, as it reminded me of normal [?] "reality"*
34. *Brought me to my senses, and therefore less in myself but more so.*
35. *The coloured fog/aura gave bulk, which helped as I kept bonking into headsets before this manifested.*

**Theme: Reflection on Death**

1. It really hit me, the lights are fading.
2. Yeah, the sadness & the loneliness when the light went out was definitely a very sudden like – sigh – loneliness.
3. And I meditate on death almost every day. But this like what happens or where do we go or where do our spirits go, I've had this vision of just like light balls communicating with each other… I definitely felt like this whole experience could have been a death experience.
4. Yeah when it faded I don't know. I think it was good. I was feeling a bit almost like alright I want to go back now. I had a bit this anxiety of coming back really. But I think the transition was effective.
5. I kind of in my head I made a story that this is how you die. But in a nice way. When you've lived your little life with your stuff, and now it's done. And this was very meditative for me. It was a perfect kind of setup to finish and now there's nothing. So the short few minutes we did of breathing was great to meditate after. So your energy builds. Then gone. cool.
6. I thought a bit about death, like my own death as well. As in like the energy that you give out to the world will always be there. In your life. Like this stuff you put out is always going to be there. And when you die, it doesn't really matter, cuz it's there, it's done. you've done it. I don't know, dispersed.
7. Yeah it definitely for me drew to my heart people I've lost in the material. And that feeling of connection and being with them.
8. [when the light faded] it did it in a gentle… it wasn't a jumping out of it. So it was… I don't know what word I'd use, but it was a gentle movement.
9. [when the light went out] I would have liked the light to continue just a little bit longer
10. I wondered where the light went, I guess.
11. I also thought it would an amazing thing to give to dying people. That would make dying a lot less scary I think… the physical body. It's just a physical body, the energy will remain, and it will stay in a connection even though this is gone. I think. It's beautiful.
12. If I was a terminal cancer patient or something I would find that a very reassuring experience.
13. Towards the end, I got rather a sad reflection about my wife's been ill. Stress breakdown five years ago. It was all to do with menopause and other things. And she hasn't ever come back as the same person. And she's become alcohol addicted. Which has been very hard. And it just alarmed me that she's 20 years younger, but she's on a sort of decline path where she hardly exercises. But when the lights were all there, there was a kind of horizon, which was a treeline. To me it was a specific place a campsite we regularly used to go to as kids up in Norfolk, stiffquay. We have tried to talk about funerals, and she wants her ashes scattered up there. So I think I was in rather a sad place for a little while. But not overwhelming. But I couldn't help seeing these trees as the horizon of the salt marshes. And I'm dreadfully afraid. She's kind of half lost the will to live you know. Her energy is just…. But recently and with my daughter coming. After awhile I'm away… this is the first time I've been away for quite awhile, 2 or 3 years, since the last conference is the last time, and she's walked the dog, and she's been shopping with the daughter, and so forth. So we're hopeful. But that was just a very personal point. that the treeline as I saw it behind the lights reminded me of it. Set me off on a sad pathway… Maybe it was the parting company with it that set me off. Cuz then looking at the horizon… sort of sad to leave it.
14. [when the lights faded] the darkness was actually really peaceful. Yeah….
15. [when the lights faded] It was like dying… It just felt like a really peaceful ideal death.
16. Yeah I could hang out in the darkness… Just there, with the slow fading [of the lights]. Now I have a sense that if I feel alone, or remote, or parted from [loved ones].. It's rest, isn't it? It's rest. All that responsibility is gone. Peace. It's very peaceful. It wasn't anxious at all. It was very peaceful.
17. Before that, before losing the individual light, when all the patterns and things disappeared, I was like 'oh my god this is what dying is like' or something like that. I felt sad then… It wasn't scary. I really really love darkness.
18. [the lights fading] felt natural to me. It was very much flowing. I don't know. Going dark it felt like it's what it has to be. I don't know.
19. It was quite bittersweet in a way. The way when we got together in the circle with our hands like that. I don't know if james was next to me or opposite me but I was kind of looking opposite as if it was James and thinking… it was like when we pass into the next life, this is what you're going to look like to me, and this how we're going to experience each other in the afterlife
20. I feel really present. It really felt like a journey. And I felt sad at the end. That that journey was at and end, so to speak. Made me feel, made me think about life. And the journey of that. Although that felt sad, that was ok. Because I felt there. And present.
21. *It made me feel that I know what loved ones in the afterlife would manifest/appear to me as.*
22. *The end felt like a peaceful death. The darkness & stillness at the end felt so peaceful.*
23. *Perfect for people who do not want to do psychedelics but have a similar experience*

24. *Dying people, accept that after leaving the physical body there is still everything (and more)*
25. *Sense of death, and knowing that the energy you have put out to the world/space will always be there.*
26. *I also felt connected to my dad who died 4 years ago – such gratitude to you all for that.*
27. *This whole experience was the liminal space between life and death, with light as the feedback loop, transferring energy to and from the field of reality. Many times I imagined us as telepathic clouds of spirits greeting one another in the after life.*
28. *As humans, we questions ourselves about what happens after death – I am afraid of this, but experiences in nature, poems, art and awe experience alleviate this fear greatly… this experience is no exception. Thank you for helping make the "unknown" a little more friendly. Thank you from the bottom of my heart.*

**Theme: Supportive Setting**

1. In a lot of VR experiences I've had you're very aware of the person standing next to you who is the controller, and he's maybe pushing you know… I don't know. This felt like because so much happened with very minimal movement you know I didn't feel the need – and you were guided through it in a very gentle way – it alleviated the stress I've previously felt in other VR situations, of like what am I doing, am I doing it right?
2. And I totally understand why you led us in blindfolded and stuff like that… It really makes you feel like you're in that space. You know the transition is so much more gentle than being aware of this person standing next to you in a hall full of people, which is normally what happens.
3. Considering this is the first time you've ever done it properly, it's amazing how harmonious it is. So carefully thought out.
4. I think it's really well curated actually.
5. *Seeking connection – feeling safe and curious in the space.*
6. *Very well conceived, designed, organised and performed. I was struck by the sensitive coming together of technology and holistic spirituality. I enjoyed sharing and trusting in the group.*
7. *I also thought the intro and aftermath was really well done.*
8. *I feel like I've been very gently guided through a journey of light, stillness and silence.*
9. *I appreciated the careful use of ritual for creating the arc of the experience, preparing for, and then closing the experience, it felt smooth, the use of music, the calm yet authoritative voice of the guide, the tips on how to explore the boundaries/possibilities of the space, the changing visuals & the sense of agency in exploring those, and the combination of a shared + also individual experience. I'm surprised at how profound it was. A good contrast to the 'headiness' of the talks.*
10. *Voice guidance was perfect*
11. *Appreciated gentleness and "heldness" of space & interactions with the people putting on gloves, headsets etc*
12. *The guiding voice and music was very soothing and calming.*
13. *Being brought in blindfolded made the world we were brought into feel way more real – the size and the air con felt like they were the real dimensions and gases floating in the room.*
14. *I have found the preparation very effective. Its ceremonial quality helped to tune into the experience. I think that was super important.*
15. *A real feeling of connectedness + compassion for the people in my group and the people also making the experience possible – almost like the feeling of being in ceremony, or breathwork, with people caring for you on your journey + healing.*
16. *I loved the music + the feminine voice, the combination of such a delicate, gentleness with the tech/masculine to heighten my empathy + feeling state– in my body (where tech can of course be linked with more masculine + the mind)*
17. *I was happy for the voice guiding us and helping us explore our environment*
18. *I am partially sighted with limited vision in one eye so I don't expect VR to be something I could access. But it was great! Really immersive. The headsets were not too heavy and fit comfortably over my eyeglasses. The gloves were light and flexible, so it was very comfortable. The voiceover, music and sound effects and cool breeze all added to the experience. I feel as if it was floating in outer space. The team handled my health consideration really well.*
19. *A very nice way of preparing me/us to get into the space and into the experience. It was a very strong potential for integration + preparation for other experiences being those therapeutic or for other altered states of consciousness.*
20. *Soundtrack was beautiful. Noise outside was distracting at start. The instructions to collaborate worked well. More guided collaborations would be great.*
21. *I enjoyed the visuals and means of interactions with them you have chosen and especially during the group tasks with working together, which everyone did in a way that was wishing to create the best experience not just for themselves but for others.*
22. *Could be more exploratory and less directive, although having direction is also good for creating the communal experience. Music beautifully curated/created. Blindfolding + journey really good. Very safe + held + fun. Good storyline.*
23. *One consequence of the limited range of phenomena was that I didn't feel rushed in experiencing things as has happened in some other VR programs where there is so much happening and one doesn't want to miss out on anything that's zipping by.*
24. *Would love to meditate a little bit more at the end, because having this switched off from real world experience made it easier at the end not to think about real world.*
25. Well I think even us being led through we kind of gained that trust, and it felt safe. So already we're in that sort of headspace. Much like a shaman whose facilitating something. Yeah I think context… it was really amazing. Thank you.

**<u>Theme: Noetic Quality</u>**

1. At one point there was strong kind of waves of emotional feeling, just kind of pulsing through the body. But It was kind of a short episode. But I wasn't expecting that actually. TBH.
2. You made me cry - I think my most profound moment was I thought they should take the VR glasses off me cuz I'm going to ruin the electrics with my tears.
3. I also quite like the bit when it just becomes red and you're no longer manipulating things and it's just an aura. Cuz the manipulating things was fun, but that was quite profound, and that's when I started to cry.
4. There was one point that I turned away from everything that was being created, and you could see the mountains in the distance and just recognizing the ironic light that all of us are in this expansive space. And yeah, it's very beautiful.
5. And this, what you've produced, was even further than that because it was this absolute truth that we are made of light. And being able to see that at that level and jump into that was profound.
6. Absolutely profound. Incredible.
7. I love the bit when you get quite excited where you can move it all around. And then there's the guidance to just be. And then all the radiation of the red, and just being and being, and that just being one of the most beautiful bits. And you could see everyone just being. That was so lovely. There were so many lovely parts. I got quite obsessed with just looking out. At the void. It's quite me. Like turning around and being like whoa.
8. [when the light went out] I had a bit of reverse kind of… I felt like I was in the matrix kind of thing. Like not there, so it was kind of a blank unconstructed space of what is, of our reality, if that makes sense. Like somewhere in between that world and this world.
9. It feels profound. Very deep.
10. In a weird sense it felt very natural… Not virtual at all… And You completely lose a sense of frontier. Like material frontier. You can choose to just get real close but not touching, merging without touching. With skin you have your atoms, you know [slapping her skin], they won't merge with each other. And you can completely choose to merge very gently and take in your energy or smoke or whatever. And it felt very natural after all, this [the real world] almost feels unnatural.
11. *wonderful to have a glimpse of what it would be like to see beyond appearances/the surface/the material.*
12. *It felt so real. More real than real at times. This has given me a felt experience of the cognitive knowing I already had of the light inside people and beings.*
13. *Seeing the essence of energy…myself…and others.*
14. *OVERWHELMING FEELING: "IT WAS MOVING THROUGH ME"; LIGHT, LIGHT HEARTEDNESS, LAUGHTER, VIBRATIONS. CONNECTION*
15. *Joy, wonderment, amazement*
16. *First sensation was a huge familiarity and I felt like crying. Felt in peace and a bit emotional in the beginning. Feeling of coming home.*
17. *Such a feeling of beauty, tenderness and connection – I leave the experience with a heightened perception of life, the love beneath, within all 'things' and the feeling of the infinite nature of 'being'.*
18. *What can I say, it was far more beautiful and inspired than I had anticipated… it really did make my heart burn with appreciation for the simple fact that I exist… I'm a biologist, a bit of a materialist, quantum physics does my head in, and I find meditating infuriating and it makes me impatient. This really did make me feel all the ideas from "sacred traditions" and various strands of philosophy that I normally never "get".*
19. *Technology and awareness held in matrimony. To generate the self "out-of-time' [can't read last word ?] and into the shared heart of selves, if just for a moment, can inspire feeling. A feeling that there is more, and that unknown is simpler than we imagine. It is our own intellectualization,[?] rationalization and judgment that gets in the way. To be [can't read this word…?] out of the way and into a new world that lives ever-present around us, is to be [can't read the word] towards a more heartful understanding of life in situ. I am grateful for the work, for its application would catalyse a much needed growth of heart and communion. Too much of secular mindfulness is individual. Shared experience is substantial.*
20. *This was a profound experience of finding out of how it is to just 'be'*

**Theme: Transcendence of Space & Time**

1. And the manipulation of space felt a little bit telekinetic in a way. I have the power!
2. I think my most profound moment was when I went to the floor, and then I could see it shooting into the floor, and I felt like I had broken into a 4d space, like this hypercube, and uhhh.. I felt like I had broken out of the moment into a parallel moment, and had double consciousness.
3. Yeah, one moment that really struck me was this idea see into your past. Because you can see the trajectory of the mark that you're leaving. And that was like an expansive moment, that realization that we're so connected to the movement and to the simple just concrete things that were happening. But that had this big significance you know like the trajectory you are leaving behind.
4. The space felt huge. I'm surprised to see how physically close we are to each other. The perception of where we were in relation to each other was vast.
5. It also made me think about time, and not being linear, and just right now, and what you leave behind you as you move, does stay behind and trail.
6. Yeah, I feel like I did. It was like the bits I could see on the floor, when I got to the floor, I was then in that pattern, and then I was down there and it was around me. So it felt like I was underneath the pattern, or in the pattern which was on the floor previously. I don't know if that's what you mean.
7. I'm annoyed at how normal this room is.
8. It feels like we really were in another world. Coming into such a big room. It's so different.
9. I think like I was quite curious how it would feel just to be in this very dark space where you can sense the space around. It's not entirely dark. There's no motion no light, nothing. And just wait in there. I felt it was quite peaceful. It was a bit… like you think is it a million years passed, or just a second? Cuz there's no stimulation, nothing happened.
10. A real sense of opening up actually. Like the space opened up around me I think. Even though it was quite focused. And that's what I was left with actually
11. Yeah it was a bit of shock seeing the screens and the lighting, and – The room was a bit surprising. I totally forget I was in a room in university.
12. Yeah one minute I thought I could actually go through [the ground]
13. You know you're in a room, but it just blankets it out in that space instead of actually in the room.
14. And I suppose we've got no preconceived idea of what the room looked like. So essentially you were just coming in to that space.
15. Yeah that *was* the space..
16. I had no idea – the room didn't feel this big. It felt like it was a small space and then you could see off into the background. It didn't feel like it was much bigger than – in fact the mat's a lot bigger than I thought the mat was.
17. Time runs differently in there.
18. *Cable slightly distracted from the feeling of being in the room*
19. *I was totally disorientated by the room when I opened my eyes.*
20. *The vastlessness [sic] of space, the non-linear time, care and gentleness to other humans in this experience, and the self-love, helped push away the torture of existential anxiety.*

**Theme: Insights to carry into everyday life**

1. It's all energy. It's all light.
2. I definitely feel an excitement and optimism for the future to have technology like this so that everyone could experience it. Like how that would change people's dynamics together as a society and how it could bring them together to work in a more harmonious way. It would be really quite something.
3. Well I hate meditating but now you make me want to try it.
4. Didn't you feel like [the light] would stay with you anyway? Like now you've got the knowledge of it, it's almost like a light has shown through. And like oh, we *are* like this, and we *do* have this light inside us. And we can look at each other, you know strangers on the bus, people, tourists that get on your way in Greenwich, and not hate them. And respect each other.
5. It's a little silly, my cat is ill at home and I've been feeding her all kinds of medicine in the past week and really worrying about her. And now I kind of feel like I should bring a mudra to her, and share my light. I was given this hippy book about natural care for cats and this stuff. Herbs and crystals for your animals. But now I'm feeling like ok maybe I'll play with this. It can't harm.
6. Yeah and just that feeling that you do radiate out. Like you were saying. You can radiate light out. And l liked the fact that it was the pressure that made it radiate. Sometimes when things are hard you need to add more pressure to be more light. It felt like that. Yeah, it felt like that.
7. [the light] is there all the time, you just can't see it.
8. Another thing I noticed is when the final gong went, I could still see the energy thing [even though the light was gone]. So I still had a closed eye visual of the thing we were playing with.
9. It really stays with you this experience I think. I don't know who was the voice, but that was also really good. With the voice – you know take it and remember.
10. …we are all the same. And sometimes we forget… just the limitations of our senses. I think I will remember this, and in difficult situations connect back to it. To this feeling of being able to create and connect.
11. This experience is stored somewhere in yourself. And if you look at someone you can recall this experience as a kind of a connection. And if you visualize that, it's a beautiful thing. In the real world. To connect.
12. She said [at the end of the narration] 'what are you left with?'. I was thinking about that. I'm left with a sense of a focus but also of an enlarged-ness. Do you know what I mean? When I go outside now I think that's what I'm left with. A kind of connectedness. But also a focus. Just that having the hands and the lights. That was like, I was like wow.
13. I think what you were saying at the end about treating everyone as equals – and what you were saying as well – that we're all, it's all just all light. That's a thing to take away isn't it.
14. And I loved it when people just became smoky…Oh the smoky beings. Oh you were so holy. Smoky light beings. That's how I'm gonna see everyone from now on.
15. So I'm gonna take this with me – the sense of – even if it's dark it's just the light, you might not see it, but it's there in a way.
16. Bit sad [when the lights faded]. But then you're internalizing it [all]. So you sort of hope that it's still there even though you can't see it.
17. *The light of the mudra will stay with me for a long while.*
18. *I will be using mudras more in my daily life.*
19. *I loved the music + the feminine voice, the combination of such a delicate, gentleness with the tech/masculine to heighten my empathy + feeling state – in my body (where tech can of course be linked with more masculine + the mind) this takes you to the feeling place which is what the world needs as we move through this paradigm shift + from one pattern to the next*

**Theme: Sense of Beauty**

1. Yeah, the sadness & the loneliness when the light went out was definitely a very sudden like – sigh – loneliness.
2. That was incredible.
3. Exquisite.
4. Holy Shit.
5. Oh my god [crying].
6. Yeah I really really liked it. It was great. it was very beautiful. Just really beautiful.
7. Beautiful voice. Beautiful voice.
8. And lovely words.
9. It's so beautiful all that light. I was trying to work out the differences, what moved when I changed mudras and what didn't.
10. Intense. Yeah. Really beautiful actually.
11. I really enjoyed the sense of wonder in this. Loving this.
12. That's really amazing.
13. Brilliant. Brilliant.
14. *That was beautiful, gentle and heart opening.*
15. *This was a wonderful and magical experience.*
16. *Beautiful experience visualising the connections*
17. *Looking up triggered a sense of awe at the size of the dark sky and the calmness of the molecules suspended in it, like a different type of atmosphere in a foreign world.*
18. *Beautiful aesthetic. Refreshing to see restraint. No need for colourful fireworks or more of an environment!*

**Theme: Comparison to Other Altered States**

1. It's kind of a bit like, it reminds me of a mescaline trip. It's almost like a pandora's box has been opened and the universe has been revealed. in its true form, rather than the virtual model that's created in our own heads. You're seeing a deeper form and fundamental truth. And the feeling that you then, once you've seen it you can't go back. [laughing]
2. It reminded me of a journey I went on at a party… and I was experimenting with a dissociative. And I didn't' mention to anyone else that I was gonna take a dissociative. And I took the dissociative. And unfortunately for the party. There was about 18, 20 people at the dinner party. And they all thought I'd gone insane. Right cuz they had no warning that I had taken anything. And I had taken such a miniscule dose. But it was because of the red wine interaction. But what happened was I went into a different dimension. I went into a $5^{th}$ dimension where everyone had these pods, and everyone who was in the party was also in the $5^{th}$ dimension with me in these pods and they were all, like 18 pods, spherical, completely incredibly technologically advanced, and every single person had everyone else's thoughts, like it was completely telepathic. Everyone knew everything.
3. My first mushroom trip, if you'll appease me with the story. I saw – I was walking around the streets of boston, I took like 3.5 grams, I thought that's how much you were supposed to take. For the first like 10 trips I did, I didn't know you could take less. So the first thing is I saw this light beam coming up from the center of everybody going up into the sky, like an umbilical cord. And every single person I passed had it.
4. Little bit. Yeah I've experienced similar realms but I've played with them less with my hands. So its normally been like very much I'm being still and it's going on around me. So in that sense this was more interactive. I've definitely been in realms full of shapes an prisms like we're flying through. So that felt familiar. But not to be playing with one's hands.
5. For me it had quite a lot of overlap with a DMT experience, where there's the sense of a reality that interpenetrates our reality, which isn't sensed directly. And that in your inner eye, in the same way as you can still see this experience if you look inwards, you can still perceive the DMT or psychedelic experience if you look inwards. It has the same quality as kind of being seen but not seen.
6. It's like my ayahuasca background. It was disconcertingly similar. It was like, is that there, or is it in my head… Lots of ayahuasca visions with lots of empty plains and mountains in the background.
7. I would like to have that visual all the time. It reminded me a lot of a particular ayahuasca experience I had, where I had like 7 hours of weeee, I'm a magician! And I was thinking this is the way it is really all the time, I just don't see all this energy coming from my hands. All the time, I have to remember that. So this was like a really great re-experiencing of that. A little bit less colorful but it was the same. And then the communal part of it like the untangling of the circle and then turning around and seeing the circle from the inside, it was like 'waaaaaaa'.
8. *Similar to aspects of Ayahuasca*
9. *The experience was similar to a 5-MEO-DMT experience in that both explore the realization of pure energy, pure being, with no reference to physical space.*
10. *It was very much like an ayahuasca experience I had once.*
11. *Similar in some ways to aya experience in the past*
12. *although it felt very strong and powerful, my psilocybin experience was beyond any of what I lived in my life in sense of awe, dissolution, joy, timelessness.*
13. *I once had a 'God trip' and voyaged to outer space on LSD. This was reminiscent.*
14. *Crossover with DMT experiences - inter-penetrating reality*
15. *It was truly transporting and illuminating, like the quiet point of an acid trip where extreme sensory experience ends and calm beauty settles in.*

**Theme: Ineffability**

1. Yeah, it certainly takes you somewhere. A different place. Yeah. It's quite hard to verbalize it. Almost like it's difficult to describe a trip. It's… It's… The words are still forming in my head.
2. Uhhhhh…. Ok. Words. Huh. I know some words… huh.
3. I can't. I don't have any words yet.
4. I felt touched by this. I thought it was very sweet. I couldn't necessarily pinpoint and label an emotion as such.
5. Yeah I've got no words.
6. *That was a wonderful unusual experience. Words seem inadequate*
7. *Feel moved, touched, quite quiet, words are hard to muster – I appreciate that because it indicates that I've been brought into my body, into my heart/energetic being.*
8. *It's hard to describe, or think of where to begin describing this experience.*
9. *I wish I could find the words to really explain more than above but my mind is pretty overwhelmed right now (in a good way!). Thank you.*

**Theme: Child-like enchantment**

1. It's like this really beautiful joyous innocent but also intimate feeling that I don't remember feeling since I was a child. Which was amazing.
2. Yeah. It's almost like that sense of awe as well, like I would get the sense of awe as a child, that it was that sort of playful innocence with interacting with others. You put up a lot of walls as you age, because of shit that's happened to you and trauma you've taken on. So to have that all dissolved, and just see it. It's quite emotional actually.
3. I like the dance of it…The sort of patterns and moving around other people's patterns. Like Sparklers…. like moving sparklers…Fizzy light… there's something very enchanting about that experience as a child.
4. I really like the childish feeling. That we were like children again, exploring and no judgement or anything. Everyone was sort of – I just felt really connected to people I didn't know. So it was just… incredible. And yeah that childlike wonder.
5. We miss out on that kind of child play in the everyday world, and it's really important. Liberating, freeing.
6. *The sense of connection and openness to the other members of the group was a very moving thing that was able to bring forth such positive emotions that were reminiscent of how one once felt as a child when interacting with other beings.*
7. *Was fascinated by the movement and development of the forms from the mudra – the different manifestation. It seems to me, at one point, that the more fluid forms began with the straight line, kind of molecule forms and then became the more fluid form. When Being was suggested, holding still, the slight movements generated little straight line shapes which I found particularly fascinating. That they were almost generated by the inevitable, slight natural movements of the hands. The red mist added a whole other element to it*

**Theme: Metaphors for day-to-day life**

1. It was profoundly lonely when everyone's light disappeared as well. It's like that cutting off, of either giving out your own light or refusing to see it in other people is kind of like a really very crystallized example of what that's like.
2. *The gradual changes from full molecules to gassy ones to swirling worms felt like a journey, some sort of evolution – an expectation about where we were going.*
3. *I was most taken with how playing with the grids seemed like we were in control, but when asked to stop the [molecular organism] did [its] own stuff. It was an acceptance of the co-creation of environment and selves. It was astonishingly calm when we relaxed at the end. It was as if a powerful step down of need to control.*
4. It's interesting, people kind of work out how it works, and then they start interacting. I guess that's how life works in general. Like oh I can do stuff, I can wave to this other person, ooohh, the light waves back, it's fascinating.
5. [not knowing the boundaries of acceptable contact] is kind of like real life.
6. I was like I would really like to move around now but I can't because there's a cord, or is there a cord? I don't really know. Maybe there's a person behind me that I don't really know about. Can I just go across the room now? So there was all these [questions], which is also kind of a spiritual [question], like the VR experience becomes sort of like, is there a cord in real life? Can I do what I want? Do I have to hold myself back now? Am I ruining it for everyone else?